\documentclass[preprint,12pt,numbers,sort&compress]{elsarticle}

\usepackage{amssymb}
\usepackage{tikz}
\usepackage{amsmath}
\usepackage{hyperref}
\usepackage{alltt}
\usepackage{slashed}

\usepackage{setspace}
\usepackage{color}
\usepackage{placeins}
\usepackage{longtable}

\usepackage{geometry}
\begin{document}

\newcommand{\mhalfo}{\frac{1}{2}}	
\newcommand{\mhalf}[1]{\frac{#1}{2}}
\newcommand{\ka}{\kappa}
\newcommand{\al}{\alpha}
\newcommand{\be}{\beta}
\newcommand{\ga}{\gamma}
\newcommand{\la}{\lambda}
\newcommand{\de}{\delta} 
\newcommand{\vp}[0]{\varphi} 
\newcommand{\vpb}[0]{\bar{\varphi}} 
\newcommand{\equ}[1]{\begin{equation} #1 \end{equation}}
\newcommand{\ba}{\begin{align}}
\newcommand{\ea}{\end{align}}
\newcommand{\eref}[1]{Eq.~(\ref{#1})}
\newcommand{\fref}[1]{Fig.~\ref{#1}}
\newcommand{\ddotp}[1]{\frac{d^d #1}{(2\pi)^d}}	
\newcommand{\nnnl}{\nonumber\\}	
\newcommand{\G}[1]{\Gamma(#1)}
\newcommand{\nq}{\nu_1}	
\newcommand{\nw}{\nu_2}	
\newcommand{\nd}{\nu_3}	
\newcommand{\dhalf}{\frac{d}{2}} 
\newcommand{\fig}[5]{\begin{figure}[#1]\centering\includegraphics[#5]{#3}\caption{#4}\label{#2}\end{figure}}
\newcommand{\tw}[1]{\texttt{#1}} 
\newcommand{\beq}{\begin{eqnarray}}
\newcommand{\eeq}{\end{eqnarray}}
\newcommand{\str}{{\rm STr}}
\newcommand{\hier}{ {\color{red}\rule{1.0\linewidth}{1pt}}}
\newcommand{\colM}[1]{{\color{blue}{#1}}}
\newcommand{\colMT}[1]{{\color{magenta}{#1}}}
\newcommand{\colD}[1]{{\color{red}{#1}}}
\newcommand{\colF}[1]{{\color{green}{#1}}}
\newcommand{\Mathematica}{\textit{Mathematica}}
\newcommand{\DoFun}{\textit{DoFun}}
\newcommand{\DoDSERGE}{\textit{DoDSERGE}}
\newcommand{\DoAE}{\textit{DoAE}}
\newcommand{\DoFR}{\textit{DoFR}}
\newcommand{\Tr}{{\rm Tr\,}}

\begin{frontmatter}

\title{CrasyDSE: A framework for solving Dyson-Schwinger equations}

\author[MQH]{Markus Q. Huber}
\ead{markus.huber@physik.tu-darmstadt.de}

\author[MM]{Mario Mitter}
\ead{mario.mitter@uni-graz.de}

\address[MQH]{Institut f\"ur Kernphysik, Technische Universit\"at Darmstadt, Schlossgartenstr. 2, 64289 Darmstadt, Germany}

\address[MM]{Institut f\"ur Physik, Karl-Franzens-Universit\"at Graz, Universit\"atsplatz 5, 8010 Graz, Austria}

\begin{abstract}
Dyson-Schwinger equations are important tools for non-perturbative analyses of quantum field theories. For example, they are very useful for investigations in quantum chromodynamics and related theories. However, sometimes progress is impeded by the complexity of the equations. Thus automatizing parts of the calculations will certainly be helpful in future investigations. In this article we present a framework for such an automatization based on a \textit{C++} code that can deal with a large number of Green functions. Since also the creation of the expressions for the integrals of the Dyson-Schwinger equations needs to be automatized, we defer this task to a \textit{Mathematica} notebook.
We illustrate the complete workflow with an example from Yang-Mills theory coupled to a fundamental scalar field that has been investigated recently. As a second example we calculate the propagators of pure Yang-Mills theory. Our code can serve as a basis for many further investigations where the equations are too complicated to tackle by hand. It also can easily be combined with \textit{DoFun}, a program for the derivation of Dyson-Schwinger equations.\footnote{The code of the program \textit{CrasyDSE} is available at \url{http://theorie.ikp.physik.tu-darmstadt.de/~mqh/CrasyDSE/}}
\end{abstract}


\end{frontmatter}

{\bf PROGRAM SUMMARY}
\begin{small}
\medskip 
\noindent \\
{\em Program Title:} CrasyDSE                                        \\
{\em Version:} 1.1.0\\
{\em Licensing provisions:} CPC non-profit use license                                  \\
{\em Programming language:}  \textit{Mathematica 8} and higher, \textit{C++}                               \\
{\em Operating system:} all on which \textit{Mathematica} and \textit{C++} are available (Windows, Unix, Mac OS)                                         
\\
{\em CPC Classification: } 11.1  General, High Energy Physics and Computing \\
    \phantom{Classi}11.4 Quantum Electrodynamics\\
    \phantom{Classi}11.5 Quantum Chromodynamics, Lattice Gauge Theory\\
    \phantom{Classi}11.6 Phenomenological and Empirical Models and Theories                      \\
\noindent
{\em Nature of problem:} Solve (large) systems of Dyson-Schwinger equations numerically.\\
{\em Solution method:} Create \textit{C++} functions in \textit{Mathematica} to be used for the numeric code in \textit{C++}. This code uses structures to handle large numbers of Green functions.\\

{\em Unusual features:} Provides a tool to convert \textit{Mathematica} expressions into \textit{C++} expressions including conversion of function names.\\
 
{\em Running time:} Depending on the complexity of the investigated system solving the equations numerically can take seconds on a desktop PC to hours on a cluster.\\
 
\end{small}

\section{Introduction}

Strongly coupled field theories play an essential role in the physical description of nature. Both established theories like quantum chromodynamics and conjectured ones like technicolor theories cannot be fully understood without non-perturbative methods. Typical approaches include Monte-Carlo simulations on a discretized space-time or functional equations. Functional renormalization group equations, see, e.~g., \cite{Berges:2000ew,Pawlowski:2005xe,Gies:2006wv,Rosten:2010vm}, Dyson-Schwinger equations, see, e.~g., \cite{Alkofer:2000wg,Alkofer:2008nt,Binosi:2009qm,Roberts:2012sv}, and the n-PI formalism, see, e.~g., \cite{Berges:2004pu}, belong to the second group. Their advantages are well appreciated and they provided many new insights.

In this article we will focus on Dyson-Schwinger equations (DSEs) and propose a concrete way to handle them when they become too complex to be treated by hand alone. DSEs consist of a system of coupled integral equations which relate different Green functions. Since there are infinitely many Green functions there are also infinitely many DSEs. Unfortunately no subset of these equations forms a closed system so that we have to deal with an infinitely large system of equations. Naturally one hopes that only a (small) finite number of Green functions is relevant and looks for truncations capturing the most important features of the theory. Of course, in order to check the validity of such an approach one should test the influence of neglected Green functions. However, this is often very tedious. On the other hand there are also theories where it is known that current truncation and approximation schemes and available methods are insufficient and need to be extended. For example, standard truncations restrict the DSEs to one-loop diagrams \cite{vonSmekal:1997vx,Alkofer:2000wg,Fischer:2003rp,Fischer:2008uz,Binosi:2009qm} but Yang-Mills theories in the maximally Abelian gauge require the inclusion of two-loop diagrams in order to be consistent in the non-perturbative regime \cite{Huber:2009wh}. Consequently the present technical methods have to be improved.

The reason why more sophisticated truncations or more complicated theories require so much more effort is mainly that the length and complexity of the explicit expressions increase considerably with the number of interactions and the number of external legs. Also the numbers of dressing functions and diagrams grow for higher Green functions. We will illustrate this below explicitly with the example of Yang-Mills theory coupled to a scalar: We will see that extending a simple truncation beyond the propagators by dynamically including the vertex between the gauge field and the scalar triples the number of dressing functions to be calculated and requires five times as many loop integrations. Furthermore, the corresponding integration kernels are substantially more complicated than the first one. Seeing such complexity arise from such a simple extension we felt it was time to think about automatizing this process. This seems even more necessary since computing time is no longer as restrictive as it was ten years ago. For example, fourteen years ago the first solution of the DSE system of Yang-Mills theory that was complete at the propagator level \cite{vonSmekal:1997vx,vonSmekal:1997is,Hauck:1998fz} relied on an angle approximation and took several hours. Nowadays it is possible to do it with the full momentum integration in a few minutes. Thus more complicated truncations and theories are definitely doable. However, right now one has to invest much time in deriving DSEs and implementing them. In a sense we fell behind the possibilities today's computers offer and we think we should try to change this and find means that allow us to focus more on the physical rather than the technical problems.

The technical part of investigating a theory numerically with DSEs consists of two main steps: First, the equations have to be derived. Second, one has to implement them in a numeric code. A tool that assists in the first part is already available with the \textit{Mathematica} \cite{Wolfram:2004} application \DoFun\ \cite{Huber:2011qr}. Here we present a generic numeric code that can serve as a basis for the second step. \textit{CrasyDSE} (Computation of Rather lArge SYstems of DSEs) is capable of dealing with a high number of Green functions and their dressing functions. Furthermore it provides several predefined integration routines and numerical approximation techniques. It can also be extended to multi-core environments (see comments in section \ref{sec:ImplementationC_quad}) and finite temperature (see comments in section \ref{sec:ImplementationC_interp}). Part of \textit{CrasyDSE} is the \textit{Mathematica} package \textit{CrasyDSE.m} to generate \textit{C++} expressions for the kernels. This first of all alleviates the generation of the code tremendously and reduces human errors and secondly is the easiest way to transform the notation of the user into the notation of \textit{CrasyDSE}. Note that the functions of the package can deal with all regular \textit{Mathematica} input and do not rely on \textit{DoFun}. In order to use the package, the file \textit{CrasyDSE.m} has to be copied from \texttt{main\_Mathematica} to a place where \textit{Mathematica} can find it. We suggest to copy it to the subdirectory \texttt{Applications} of the \textit{Mathematica} user directory (\texttt{\$UserBaseDirectory/Applications})\footnote{On a Unix machine this is typically \texttt{\textasciitilde/.Mathematica/Applications}, on Windows it is \texttt{User\textbackslash Application Data\textbackslash Mathematica\textbackslash Applications}.}. Now the package can be loaded with \texttt{<<CrasyDSE`}.

In the following we will describe the general procedure to solve DSEs in section~\ref{sec:solveDSEs}. The numerical problem is formulated in section~\ref{sec:Problem} and section~\ref{sec:ImplementationC} contains details on the provided routines to solve DSEs. Secs.~\ref{sec:example} and \ref{sec:YM} explain the application of \textit{CrasyDSE} using as examples the calculation of two DSEs of a scalar field coupled to Yang-Mills theory and the calculation of the propagators of pure Yang-Mills theory. Finally, we give a summary and an outlook in section~\ref{sec:outlook}. In three appendices we provide details and summaries on functions and variables of \textit{CrasyDSE}.

\section{Solving Dyson-Schwinger equations}
\label{sec:solveDSEs}

\begin{figure}[tb]
\begin{center}
 \includegraphics[width=\linewidth]{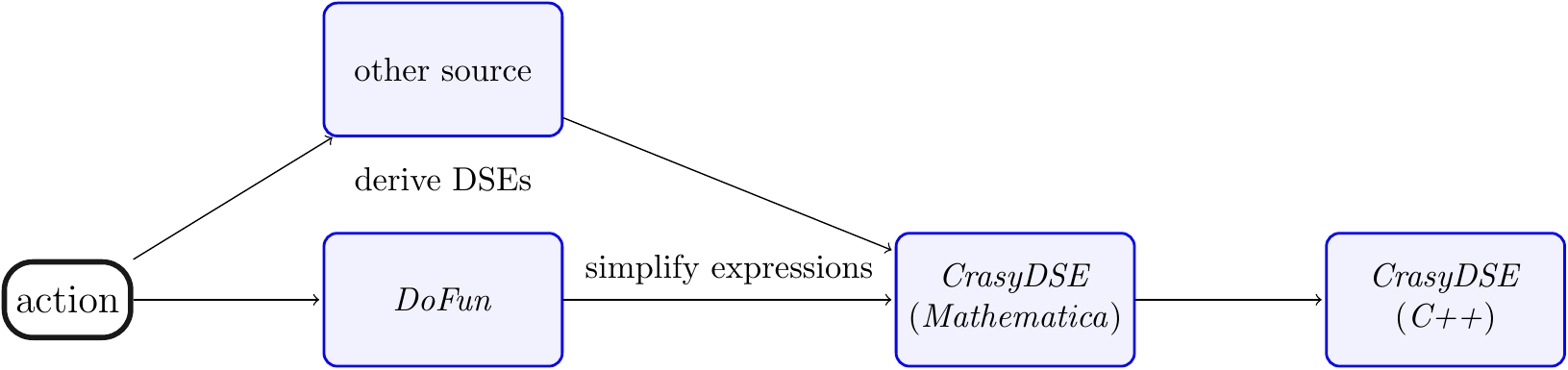}
\caption{\label{fig:prog}Schematic workflow.}
\end{center}
\end{figure}

Our approach to solving DSEs can be separated into three parts as illustrated in \fref{fig:prog}: 
\begin{enumerate}
 \item Derive the equations from the given action by the method of choice. If not done in \textit{Mathematica}, enter them into \textit{Mathematica}.
 \item Use the \textit{Mathematica} package \textit{CrasyDSE.m} to generate the \textit{C++} files with the kernels. Alternatively, in simple cases one can write the kernel files manually.
 \item Use the kernel files with the \textit{C++} code of \textit{CrasyDSE} to solve the DSE numerically.
\end{enumerate}
We illustrate the last steps with two examples in sections \ref{sec:example} and \ref{sec:YM}.

For the first step, the derivation of the DSEs, we recommend the \textit{Mathematica} application \textit{DoFun} (Derivation Of FUNctional equations) \cite{Huber:2011qr}. Its predecessor is the \textit{Mathematica} package \textit{DoDSE} (Derivation Of DSEs) \cite{Alkofer:2008nt,Huber:2012th} which was of great help in the investigation of big systems of DSEs like that of the maximally Abelian gauge \cite{Huber:2009wh} and the Gribov-Zwanziger action \cite{Huber:2009tx,Huber:2010cq}. The calculation of some infrared properties in the maximally Abelian gauge would even have been impossible without automatization due to the huge number of terms \cite{Huber:2011fw,Huber:2012th}.
Later on \textit{DoDSE} was considerably extended and the derivation of functional renormalization group equations\footnote{Recently a similar program to \textit{CrasyDSE} has become available for functional renormalization group equations with the program \textit{FlowPy} \cite{Fischbacher:2012ib}.} was included \cite{Huber:2011qr}, thus it was renamed to \textit{DoFun}. However, note that \textit{CrasyDSE} works completely independent of \textit{DoFun}.

The second step consists in making the \textit{Mathematica} expressions accessible for \textit{C++}. We chose to write our own functions that generate complete \textit{C++} files. Thus we  maintain as much control as possible over the process and make it more transparent for the user. However, in principle it would also be possible to let \textit{Mathematica} and \textit{C++} interact in a more direct way via \textit{MathLink}. All the necessary functions to create the \textit{C++} files are included in the package \textit{CrasyDSE.m}, but the notebook from which it is created is also provided so that the user has direct access and can most easily adapt code if required.

Finally, after all kernels have been written into \textit{C++} files, one uses the provided \textit{C++} modules to solve the equations. Typical initial work includes defining model parameters, defining the required Green functions and their dressing functions, choosing integration routines and defining the renormalization procedures. The way dressing functions are defined is thereby quite arbitrary: One can use closed expressions, for example, from fits to lattice data, interpolations or expansions in sets of polynomials.
In order to help with starting calculations with \textit{CrasyDSE} we provide several examples with the main code. 

We want to stress that \textit{CrasyDSE} can not be considered a black box. In order to successfully use it, the user has to understand many of the employed routines and adapt them if required. \textit{CrasyDSE} is merely a framework for solving DSEs that provides structures to handle Green functions and their DSEs and modules to perform the most basic steps like integration. However, as every problem has its own intricacies the user still has to implement many specific functions, e.~g., extrapolation functions for dressing functions.

\section{General formulation of the numerical problem}
\label{sec:Problem}

As already mentioned DSEs form an infinitely large set of equations. For numeric calculations we take a subset of these equations, but they will always depend on Green functions whose DSEs are not part of the truncation. Depending on the details of our truncation scheme we can either provide expressions for them as external input or drop diagrams containing such Green functions. Furthermore, every Green function can consist of several dressing functions and before we can do any calculation we have to project the DSEs such that we deal with scalar integrals. Sometimes it is not possible or not feasible to project directly onto the dressing functions and additionally a linear system of equations has to be solved to get results for them.
But let us for now assume for  simplicity that it is possible to project directly onto the dressing functions.

Every Green function depends on several momenta. The dressing functions, however, depend only on a reduced number of variables. For example, a two-point function depends on two external momenta. Momentum conservation reduces these to one momentum. The dressing function(s) of such a Green function depend only on the remaining momentum squared, i.~e., one variable instead of four. For a three-point function there are two independent momenta  and the dressing functions depend on three variables. In a slight abuse of language we call the variables of the dressing functions external momenta. The number of variables is denoted as the dimension of external momenta. Later we will also encounter internal momenta. These are the remaining loop momentum variables after trivial integrations have been performed.

In the following dressing functions are denoted by $A^{g,i}(x_g)$, where $g$ denotes the Green function to which it belongs and $i$ labels the dressing functions of a Green function. $x_g\in \Omega_g$ represents the external momenta.
We have to solve the following integral equations:
\begin{eqnarray}\label{eq:integral_equation}
 A^{g,i}(x_g) & = & A_{\text{bare}}^{g,i} + \sum\limits_{l}Z^{g,i,l}\int\limits_{\mathbb{R}^{d_l}}d^{d_l}y F_l\left(y,x_g,\left\{A\right\},\left\{A_{\text{model}}\right\}\right)\ ,\\
  A^{g,i}:& &\Omega_g \subseteq \mathbb{R}^{d_g}\rightarrow\mathbb{R}\ ,\nonumber
\end{eqnarray}
where $A_{\text{bare}}^{g,i}$ are the bare dressing functions (if non-zero and possibly including a renormalization constant).
The sum over $l$ denotes contributions from different graphs and $Z^{g,i,l}$ are the renormalization constants of the bare $n$-point function of a given graph.
The integration is over the loop momenta $y$. The dimensions of external and internal momenta are indicated by $d_g$ and $d_l$, respectively.
The $F_l\left(y,x_g,\left\{A\right\},\left\{A_{\text{model}}\right\}\right)$ denote the kernels of the integrals. They depend on the internal and external momenta explicitly and via several Green functions also implicitly. Some of them, the $\{A\}$, are a dynamic part of the truncation, whereas others, the $A_{\text{model}}$, are given by external input.

Before solving this system of equations numerically the following steps are required:
\begin{enumerate}
  \item The dressing functions $A^{g,i}$ have to be approximated, e.~g., by discretization of the argument or expansion in an orthogonal set of functions (see section \ref{sec:ImplementationC_interp}).
  \item If the integrals are divergent, one needs a regularization prescription, e.~g., a sharp cutoff in $|y|$ or BPHZ \cite{Bogoliubov:1957gp,Zimmermann:1969jj,Hepp:1966eg} (see section \ref{sec:YM+scalar}).
  \item Expressions for the dressings $A_{\text{model}}$ need to be provided (see sections \ref{sec:ImplementationC_sol} and \ref{sec:YM+scalar}).
  \item A renormalization procedure needs to be defined to fix the renormalization constants $Z^{g,i,l}$ (see sections \ref{sec:ImplementationC_sol} and \ref{sec:YM+scalar} and \ref{sec:app_extra_user-funcs}).
\end{enumerate}
After this is settled, the integrals can be evaluated with the provided quadratures (see section \ref{sec:ImplementationC_quad}). A solution can be found, for example, 
by fixed point iteration or Newton's method, see, e.~g., \cite{Bloch:1995dd,Atkinson:1997tu,Maas:2005xh}. Both these solution methods are implemented in \textit{CrasyDSE}, see section \ref{sec:ImplementationC_sol}.

\section{Implementation in \textit{C++}}
\label{sec:ImplementationC}

Solutions to the different problems stated in section \ref{sec:Problem} are implemented
in \textit{C++} in three modules:

\begin{itemize}
  \item \emph{dressing.cpp/hpp},
  \item \emph{quadrature.cpp/hpp},
  \item \emph{DSE.cpp/hpp},
\end{itemize}
where \emph{dressing.cpp/hpp} uses the structure \texttt{dse} from \emph{DSE.cpp/hpp}.

Additionally some simple general functions are stored in \emph{function.cpp/hpp}. All
these files can be found in the directory \texttt{main}. The provided examples are located in separate directories in \texttt{examples}. They are initialized
and called in the files \emph{sphere\_main.cpp}, \emph{interp\_main.cpp}, \emph{scalar\_main.cpp} and \textit{YM4d\_main.cpp}. A very basic 
\emph{Makefile} for \textit{Unix} using the \textit{g++} GNU compiler is provided with each of the examples.

The provided modules are as self-contained as possible and can be arbitrarily extended, e.~g., by including
further interpolations in \textit{dressing.cpp/hpp}, adaptive integration 
algorithms in \textit{quadrature.cpp/hpp} or additional solving strategies in 
\textit{DSE.cpp/hpp}.

\subsection{Approximation (dressing.cpp/hpp)}
\label{sec:ImplementationC_interp}

All functions and parameters relevant for approximating the dressing functions are referenced to or stored in the structure \texttt{struct dse} defined in \emph{DSE.hpp}. A summary together with the corresponding objects in a DSE is given in table~\ref{tab:approx}.

The expansion coefficients for the \texttt{int dim\_A} dressing functions are contained in the array \texttt{double *A}. We have \texttt{int dim\_x} external variables, and the numbers of expansion coefficients for each of them are saved in the array \texttt{int *n\_A}. In the case of linear interpolation the interpolation points have to be stored in the array \texttt{double *x\_A}. The total number of expansion coefficients is \texttt{int ntot\_A}$:=\prod\limits_{i=0}^{\texttt{dim\_x-1}}\texttt{n\_A[i]}$. Since future extensions for non-zero temperature calculations require also discrete variables, i.~e., Matsubara frequencies, part of the variables can be considered as integer numbers $z$ of dimension \texttt{int dim\_mat}. Currently non-zero temperature is not fully implemented and thus \texttt{dim\_mat} should always be set to zero. The allocation/deallocation of the arrays \texttt{A} and \texttt{x\_A} is done with \texttt{void init\_A\_xA(void *dse\_param)}/\texttt{void dealloc\_A\_xA(void *dse\_param)}\footnote{In order not to overload the text we refrain in most cases from a detailed explanation of all arguments and only give them for reference. Details on the meaning of the arguments can be found directly in the code where all of them are explained in the function descriptions.}. A minimal \texttt{dse} structure can be initialized with \texttt{void init\_default\_dse(void *dse\_param)}. More detailed information on \texttt{A} and \texttt{x\_A} can be found in \ref{sec:x_A-A}.

In \textit{dressing.cpp/hpp} two ways to express the dressing functions are provided:
\begin{itemize}
 \item linear interpolation (\texttt{Lin\_gen\_dress\_interp}),
 \item expansion in Chebyshev polynomials (\texttt{Cheb\_gen\_dress\_interp}).
\end{itemize}
In the case of a linear interpolation \texttt{A} contains the function values on a rectilinear grid
stored (together with the discrete arguments) in \texttt{x\_A} whereas in the case of an expansion in 
Chebyshev polynomials the expansion coefficients are stored in \texttt{A} and only the discrete
arguments need to be stored in \texttt{x\_A}\footnote{Note that when using the
DSE solving routines also for Chebyshev interpolation the continuous external momenta have to be stored in \texttt{x\_A} using the function \texttt{Cheb\_init\_cont\_xA} described in \ref{sec:x_A}.}. 
For a Chebyshev expansion, as introduced for dressing functions of Green functions in \cite{Bloch:1995dd},
it is possible to either transform the standard interval $[-1,1]$ linearly or logarithmically
to the actual domain of interpolation via setting \texttt{int cheb\_trafo[i]} to \texttt{1} or \texttt{2}, respectively, where \texttt{i} denotes the external variable. By default it is set to \texttt{1} in \texttt{init\_A\_xA}. Furthermore, with \texttt{int cheb\_func\_trafo[i]} one can expand the logarithm of a function \cite{Bloch:1995dd} instead of the function itself by changing its default value \texttt{0}, set in \texttt{init\_A\_xA}, to \texttt{1}. \texttt{i} denotes here the dressing function. For an example see section \ref{sec:YM_ren_sol}.

The provided interpolation algorithms work only as long as the arguments $x$ of the dressing function are inside the user-defined domain $\Omega_g$. To determine if $x\in \Omega_g$ the user-defined function \texttt{void def\_domain(double *x, void *dse\_param)} is called. For details of how to construct this function see \ref{sec:app_extra_user-funcs}. If $x\notin \Omega_g$, the user has to provide a function for extrapolation via \texttt{double interp\_offdomain(int *pos, double *x, int iA, void *dse\_param)}. Details can again be found in \ref{sec:app_extra_user-funcs}, but the gist is that the array \texttt{pos} knows on which side of the allowed interval $\left[a_i,b_i\right]$ the external variable $x_i$ lies: If $x_i<a_i$ or $x_i>b_i$, \texttt{pos[i]} is \texttt{1} or \texttt{2}, respectively.

The correct initialization and definition of the required parameters and functions is illustrated
by the example 
\emph{interp\_main.cpp} interpolating three functions linearly and with Chebyshev polynomials.
These functions have two discrete and 
three continuous variables where the domains of the continuous variables depend on the values of 
the discrete variables.

\subsection{Integration (quadrature.cpp/hpp)}
\label{sec:ImplementationC_quad}

All functions and parameters relevant for integration are referenced to or stored in the structure \texttt{struct quad}. It is defined in \emph{quadrature.hpp} and allocation and deallocation is done with \texttt{void init\_quad(void *quad\_param)} and \texttt{void dealloc\_quad(void *quad\_param)}, respectively. An overview of the members of \texttt{quad} relevant to the user and their usage in the context of DSEs is given in table~\ref{tab:quad}.

This module provides the means to integrate \texttt{nint} integrals of dimension \texttt{dim}. Any integral is split into three parts: a constant factor independent of any variables, a Jacobian and the remaining integrand. These three parts have to be given by the three functions \texttt{void coeff(double *coefficients, void *int\_param)}, \texttt{double} \texttt{jacob(double *x)} and \texttt{void} \texttt{int\-egrand(double *erg, double *x,} \texttt{void *int\_param)}. They have to be defined by the user, but \texttt{integrand} and \texttt{coeff} are usually created with the \textit{Crasy\-DSE} \textit{Mathematica} notebook. Further details on these functions can be found in \ref{sec:app_extra_user-funcs}.

The boundaries of the integrals are defined in the function \texttt{void} \texttt{bound\-ary(double *bound, double *x, int idim, void *int\_param)}. Details on its required contents are provided in \ref{sec:app_extra_user-funcs}. For now it suffices to say that it defines the domains $[a_i, b_i]$ of the integration variables $y_i$, where the inner integration boundaries may depend on the outer integration variables:
\begin{eqnarray}\label{eq:domain}
 \int\limits_{a_0}^{b_0}dy_0\int\limits_{a_1(y_0)}^{b_1(y_0)}dy_1\cdots\int\limits_{a_{\texttt{dim}-1}(y_0,\dots,y_{\texttt{dim}-2})}^{b_{\texttt{dim}-1}(y_0,\dots,y_{\texttt{dim}-2})}dy_{\texttt{dim}-1}.
\end{eqnarray}
Boundaries can also depend on the external momenta. However, in this case the integration routine becomes slower. If this is required one has to set \texttt{int bound\_type} to \texttt{1}. The default value, set in \texttt{init\_quad}, is \texttt{0}, i.~e., the boundaries must not depend on the external momenta. As an example where the boundaries depend on the external momenta one can consider the integration of the Yang-Mills system in section~\ref{sec:YM}.

\begin{figure}[tb]
 \begin{center}
    \includegraphics[width=0.9\textwidth]{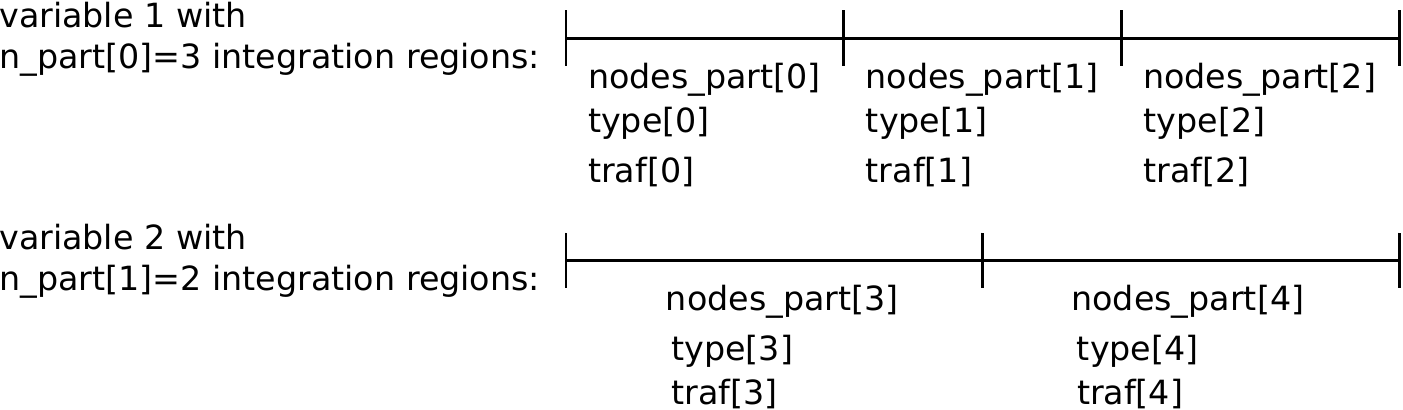}
    \caption{\label{fig:integration_regions}Example of how integration regions and related variables are used: For two integration variables the array \texttt{n\_part} contains the information how many integration regions there are for each variable. The resulting integration regions are then numbered consecutively. The number of integration points, the integration type and the transformation type of each region are stored in the arrays \texttt{nodes\_part}, \texttt{type} and \texttt{traf}.}
 \end{center}
\end{figure}

The integration of every variable $y_i$ can be split into several parts which may increase the precision of the results, see, e.~g., \cite{Bloch:1995dd}. The number of integration regions for the \texttt{i}-th integration variable is given by \texttt{int n\_part[i]}. Each region is labeled here by \texttt{j} and the number of its integration points is set by \texttt{int nodes\_part[j]}.
Fig. \ref{fig:integration_regions} illustrates the numbering of integration regions.
For each region a quadrature rule has to be chosen via setting \texttt{int type[j]} to a number corresponding to one of the quadrature rules given in table~\ref{tab:quads}, where also necessary parameters are indicated.
The quadrature rules are defined on the interval $[-1,1]$, which can be transformed with various functions to the actual integration interval defined in \texttt{boundary}. This works by setting \texttt{int traf[j]} to one of the values indicated in table~\ref{tab:trafs}.

\begin{table}[tb]
 \begin{center}
  \begin{tabular}{|l|l|l|}
   \hline
   quadrature rule & \texttt{type[j]} & parameters\\
   \hline
   \hline
   Gauss-Legendre & \texttt{0} & none\\
   \hline
   Gauss-Chebyshev type two & \texttt{1} & none\\
   \hline
   Fejers second rule & \texttt{2} & none\\
   \hline
   Double exponential & \texttt{3} & \texttt{int param[0]}: stepsize\\
   \hline
  \end{tabular}
  \caption{\label{tab:quads}Currently implemented quadrature rules, the corresponding values of \texttt{type[j]} and the required parameters. Details can be found in \textit{quadrature.cpp}.}
 \end{center}
\end{table}

\begin{table}[tb]
 \begin{center}
  \begin{tabular}{|l|l|}
   \hline
   transformation rule & \texttt{traf[j]}\\
   \hline
   \hline
   none & \texttt{0}\\
   \hline
   linear & \texttt{1}\\
   \hline
   logarithmic & \texttt{2}\\
   \hline
   modified logarithmic & \texttt{3}\\
   \hline
   modified logarithmic \cite{Horvatic:2011pc} & \texttt{4}\\
   \hline
  \end{tabular}
  \caption{\label{tab:trafs}Currently implemented transformation rules from $[-1,1]$ to the actual integration interval and the corresponding values for \texttt{traf[j]}. Details on these transformations can be found in the function \texttt{nw\_trafo} of \textit{quadrature.cpp}.}
 \end{center}
\end{table}

Finally we want to draw attention to the fact that the integration is the most costly part of solving DSEs. Therefore a parallelization of the program is most efficient in the function \texttt{void integrate(double *erg, void *quad\_param, void *int\_param)} of the quadrature module, where the loop over the internal or external momenta can be distributed to several cores. This is not implemented but a user familiar with parallelization should be able to extend the program in this direction without too much effort.

To summarize the user has to provide the following functions:
\begin{itemize}
 \item \texttt{integrand}:
	Defines the integrand of the integral. Usually this will be the kernel function created by the \textit{Mathematica} notebook.
 \item \texttt{coeff}:
	Constant factor. Usually it is created by the \textit{Mathematica} notebook.
 \item \texttt{jacob}:
	Defines the Jacobian of the integral measure.
 \item \texttt{boundary}:
	Initializes the boundaries $a_0$, $b_0$, \ldots, $a_{dim-1}(y_0,\ldots,y_{dim-2})$ and $b_{dim-1}(y_0,\dots,y_{dim-2})$, where each boundary can depend on previous integration variables.
\end{itemize}

The correct initialization and definition of the needed parameters and functions is illustrated
by a simple example.  In \emph{sphere\_main.cpp} the function
\begin{eqnarray}\label{eq:sphere_integrand}
 \texttt{sphere\_integrand}:\mathbb{R}^{3} & \rightarrow & \mathbb{R}^{3}\ ,\\
			 (x,y,z) & \mapsto & \left(1, x^2+y^2, \left(x^2+y^2\right)^2\right)\ ,\nonumber
\end{eqnarray}
times the Jacobian $r$ is integrated over 
\begin{eqnarray}\label{eq:sphere_domain}
 \int\limits_{-R}^{R}dz\int\limits_{\pi}^{2\pi}d\phi\int\limits_{0}^{\sqrt{R^2-z^2}}dr.
\end{eqnarray}
Furthermore, we exemplify here the use of external parameters. For solving DSEs the external parameters are normally the external momenta. However, the integration routine can handle also other cases of external parameters. In general one can define \texttt{int nint\_para} different parameters initialized in \texttt{void init\_para(int i, void *int\_param)} for which the integral is performed by calling the integration routine once. For the standard application in DSEs these functions are set automatically as required by the function \texttt{init\_A\_xA}. In the example \emph{sphere\_main.cpp} another possibility is demonstrated. Here \texttt{init\_para} is set to \texttt{sphere\_init\_para} which sets the external parameters as multiplicative factors for the integrals.

\subsection{Solving DSEs (DSE.cpp/hpp)}
\label{sec:ImplementationC_sol}

Assuming that together with the quadrature a proper regularization has been chosen all the
integrals in \eref{eq:integral_equation} are known and we are
left with the task of solving the given integral equations including a proper renormalization.

\begin{figure}[tb]
\begin{center}
 \includegraphics[width=\linewidth]{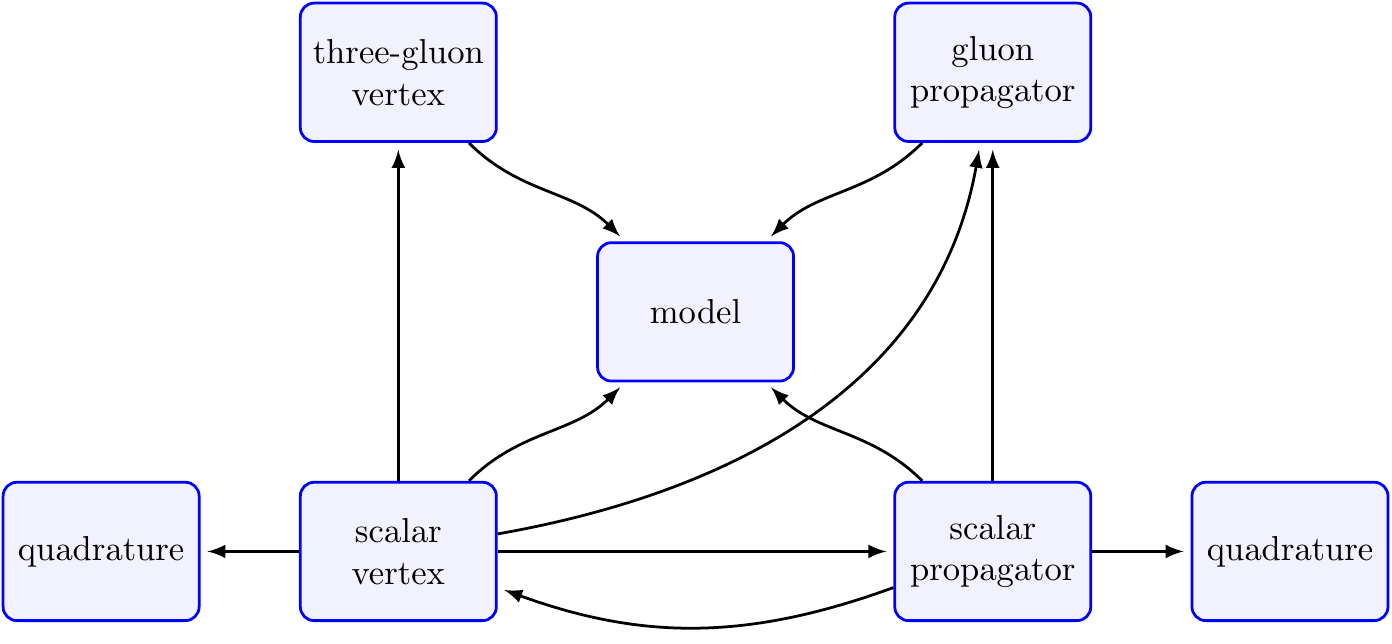}
\caption{\label{fig:structures}Schematic illustration of structures defined in the code and their dependencies for the example of Yang-Mills theory coupled to a scalar field considered in section~\ref{sec:example}.}
\end{center}
\end{figure}

Every Green function present in our truncated set of DSEs is
represented by its own structure \texttt{dse} which contains all the necessary functions and
parameters in order to evaluate its dressing functions via the function \texttt{dress}.
This is already all one needs to initialize the modeled Green functions, whereas 
those we are going to solve for need additional information in their structures.
A summary of all variables and functions relevant for the user is given in table~\ref{tab:dse}.
The fact that the equations are coupled and the iteration of one dressing function needs information
about dressing functions of \texttt{int n\_otherGF} other Green functions is handled by \texttt{struct dse *otherGF} which contains copies of
the needed structures, called when evaluating the integration kernels. Therefore it is necessary
to allocate the arrays in all other Green functions before copying them to \texttt{otherGF} such that the
correct pointer addresses are available. Only variables which are not supposed to be changed during the
iteration procedure will be copied by value.
Additionally some model parameters might be needed by the (modeled)
dressing functions which are pointed to by every structure \texttt{dse} via \texttt{struct mod},
defined by the user.
For an example see \fref{fig:structures}, where also the \texttt{quad} structures are indicated which contain 
specifics on the integration routines needed to evaluate the loop integrals in the graphs. 

Focusing now on one specific Green function 
these graphs are grouped into \texttt{int n\_looporder} contributions where all \texttt{int n\_loop[i]} 
members of one of the groups can be evaluated using the same quadrature \texttt{struct quad Q[i]}. 
The evaluation of the self-energy for the \texttt{int dim\_A} dressing functions and  
\texttt{int ntot\_A} different array points $x_g$ is then performed by
calling \texttt{void selfenergy(void *dse\_param)}. The result is stored in \texttt{double *self\_A}. It will be used in the user-defined function \texttt{void renorm(void *DSE)} to obtain the new dressing functions, see \ref{sec:app_extra_user-funcs} for details.

\subsubsection{Iteration}
\label{sec:ImplementationC_sol_iter}

We implemented a fixed point iteration solution technique, i.~e., the system is solved by calculating
\begin{eqnarray}\label{eq:integral_equation_iteration}
 A^{g,i}_{(k+1)}(x_g) & = & A_{(k),\text{bare}}^{g,i} + \sum\limits_{l}Z^{g,i,l}_{(k)}\int\limits_{\mathbb{R}^{d_l}}d^{d_l}yF_l\left(y,x_g,\left\{A_{(k)}\right\},\left\{A_{\text{model}}\right\}\right)\ ,
\end{eqnarray}
from the previous dressing functions $A^{g,i}_{(k)}$. The calculation of the new dressing functions $A^{g,i}_{(k+1)}$
is performed in several steps, where it might be necessary that a subset of Green functions is iterated
till convergence for every ``meta-iteration'' of the full set of equations. 

Before starting the iterations the initial dressings $A^{g,i}_{(0)}$ of every Green function 
have to be set via \texttt{void init\_dress(void *dse\_param)}. 
The last step in every iteration is then to renormalize such that the $A^{g,i}_{(k+1)}$ have the correct values/derivatives
at the renormalization scale(s) $\mu$ via \texttt{int renorm\_n\_param} parameters which may need some initialization in \texttt{void init\_renormparam(void *dse\_param)}.
In the example of section \ref{sec:example} 
renormalization
will be done by fixing the \texttt{int renorm\_n\_Z} 
renormalization constants $Z^{g,i,l}_{(k)}$ accordingly in the function \texttt{void renorm(void *DSE)}. Note that renormalization constants also appear in the bare Green functions $A_{(k),\text{bare}}^{g,i}$, but one could employ subtracted equations to drop them.

The iteration of a single Green function is performed in \texttt{void solve\_iter(void *dse\_param, int output)} where different stopping criteria
(absolute difference \texttt{double epsabs}, relative difference \texttt{double epsrel} and maximum number of iterations \texttt{int maxiter})
are available. 
Several Green functions can be united in an array which can be passed to \texttt{void meta\_solve\_iter(struct dse *DSE, int ndse, double epsabsstop, double epsrelstop, int iterstop, int output)}. It has the same stopping criteria as \texttt{solve\_iter} and in every meta-iteration \texttt{solve\_iter} is called for every Green function.
This allows different relative iterations, e.~g., all Green functions are iterated once for every meta-iteration
or a subset of Green functions is iterated till convergence while another subset is iterated only once.

To get an idea of the efficiency of our code we compared the calculation from section \ref{sec:example} with an independently created code that was optimized for this problem \cite{Alkofer:2011up}. 
In general the difference in time depends on how much optimization is possible in the given problem. In the present case the time difference for one iteration was less than a factor of $2$.

\subsubsection{Newton's method}
\label{sec:Newton}

Another method for solving DSEs is based on Newton's method to solve a non-linear system of equations. It was already used in many DSE calculations, see, for instance, \cite{Bloch:1995dd,Atkinson:1997tu,Maas:2005xh,Fischer:2002hn}. For this method the system of DSEs is rewritten into the following form:
\begin{align}\label{eq:E}
 E^{(i,k)} & = - A^{i}(x_k)  + A_{\text{bare}}^{i}(x_k) + \sum\limits_{l}Z^{i,l}\int\limits_{\mathbb{R}^{d_l}}d^{d_l}y F_l\left(y,x_k,\left\{A\right\},\left\{A_{\text{model}}\right\}\right),
\end{align}
where $i$ labels the dressing functions of all DSEs and $x_k$ denotes the external momenta.
We assume here that the dressing functions $A^{i}(x_k)$ are expanded in a set of basis functions. The corresponding expansion coefficients are the unknown variables $c^{(i,j)}$, where $j$ labels the polynomials. The goal is to find those values for $c^{(i,j)}$ that make all $E^{(i,k)}$ vanish. Newton's method yields new coefficients by the following formula:
\begin{align}
 c'^{(i,j)} & = c^{(i,j)} - \lambda\,\sum_{i',k} \left(J^{(i,j)}_{(i',k)}\right)^{-1} E^{(i',k)},
\end{align}
where the Jacobian $J$ is given by
\begin{align}
 J^{(i,j)}_{(i',k)}:=\frac{\partial E^{(i',k)}}{\partial c^{(i,j)}}.
\end{align}

The backtracking parameter $\lambda$ can be used to optimize this step by choosing an appropriate value between zero and one. The determination of $\lambda$ can be subject of sophisticated algorithms, see, for example, \cite{Maas:2005xh}. Here, however, we simply cut $\lambda$ in half if the norm of the new $E^{(i,k)}$ is not smaller than that of the old one. If the starting functions are well chosen this is sufficient for the example of section \ref{sec:YM}. This procedure is repeated until the norm of the vector $E^{(i,k)}$ drops below a given value or a maximal number of iterations is reached. A single iteration step takes here much longer than for the fixed point iteration described in section \ref{sec:ImplementationC_sol_iter} because the calculation of the Jacobian is rather expensive. In principle the derivatives required to get $J$ can be done directly, but here Broyden's method is used which defines an approximate Jacobian by a simple forward differentiation with small step size $h$:
\begin{align}
 J^{(i,j)}_{(i',k),\text{approx}}:=\frac{E^{(i',k)}(c^{(i,j)}+h)-E^{(i',k)}(c^{(i,j)})}{h},
\end{align}
where the notation $E^{(i',k)}(c^{(i,j)}+h)$ means that $E^{(i',k)}$ is calculated with the coefficient $c^{(i,j)}$ changed by $h$, whereas $E^{(i',k)}(c^{(i,j)})$ refers to the original $E^{(i',k)}$.
This prescription proved very reliable for the example treated in section \ref{sec:YM}.

Newton's method is implemented in the function \texttt{void} \texttt{solve\_iter\-\_secant(struct} \texttt{dse} \texttt{*DSE,} \texttt{int} \texttt{ndse,} \texttt{int} \texttt{maxiter, double eps\_E, int} \texttt{output)}. As arguments it takes the array of \texttt{dse}s \texttt{DSE}, the length of this array \texttt{ndse}, the maximal number of iterations \texttt{maxiter}, the stopping value for $\sum_{i,k} E^{(i,k)}$ \texttt{eps\_E} and an integer number \texttt{output} which determines if intermediary output should be printed to the screen ($1$) or not ($0$). Of course Newton's method can be combined with the fixed point iteration. For example, one could solve two of three DSEs with Newton's method and then iterate the third one with \texttt{solve\_iter}.

\section{Solving the gap and vertex equations of Yang-Mills theory coupled to a scalar field}
\label{sec:example}

In this section we describe how to solve a truncated set of DSEs of Yang-Mills theory coupled to a scalar field. We will first give a short overview of the employed truncation and then explain how to solve it with \textit{CrasyDSE}.

The derivation of the DSEs is not discussed here, but we provide details in the \textit{Mathematica} notebook \textit{DoFun\_YM+Scalar.nb}.
The results of this notebook form the basis on which we create the functions for the \textit{C++} code. The corresponding steps are contained in a second notebook, \textit{CrasyDSE\_YM+scalar.nb}. It describes how the expressions of the DSEs have to be modified so they can be used as input for \textit{CrasyDSE}. In a second part all required definitions are provided and the kernels are created. We will explain only the latter here, since the first steps consist only of standard \textit{Mathematica} transformations and are not special to \textit{CrasyDSE}.
Finally we explain some details for this specific example in the \textit{C++} code. The provided files allow the interested reader to follow the complete procedure, from the derivation of the DSEs to their numeric solution, in detail.

\subsection{Yang-Mills theory coupled to a scalar field}
\label{sec:YM+scalar}

In nature elementary matter fields are fermions. In quantum chromodynamics, for example, these are the quarks which interact via gluons. However, since their spin is $1/2$, quarks are Dirac fields and consequently represented by spinors. An advantage of functional methods is that they do not suffer from fundamental problems when dealing with anti-commuting fields. However, calculations are complicated by the Dirac structure, since it allows more dressing functions than for a simple scalar, see, e.~g., \cite{Ball:1980ay}. While at the level of propagators this is still doable, see, e.~g., \cite{Maris:1997tm,Maris:1999nt,Alkofer:2002bp,Fischer:2009gk}, three-point functions become already quite tedious \cite{Alkofer:2008tt}.
Since some non-perturbative phenomena like confinement may not depend on the fields being spinors or scalars, one can alleviate calculations by replacing the quarks by scalar fields. In order to mimic quarks such fields also have to be in the fundamental representation. The calculation used as an example for the presentation of \textit{CrasyDSE} in this section is motivated by investigations along these lines  \cite{Fister:2010yw,Alkofer:2010tq,Macher:2011ys,Hopfer:2011dt,Alkofer:2011up,Maas:2011yx}. 

The renormalized action of this theory in Landau gauge reads in momentum space
\begin{align}\label{eq:scalaction}
 S[A,\bar{\varphi},\varphi]&=\int \frac{d^dq}{(2\pi)^d} \Bigg(\frac{1}{2}Z_3 A_\mu^a(q)\left( q^2 g_{\mu\nu}- q_\mu q_\nu \right) A_\nu^a(-q)\nnnl
 &+\hat{Z}_3 \bar{\varphi}^i(q^2+Z_m m^2) \varphi^i \Bigg)\nnnl
+&\int \frac{d^dq_1d^dq_2}{(2\pi)^{2d}}\Bigg(
i\, Z_1\,g\,f^{abc}q_{1\mu}A_\nu^a(q_1)A_\mu^b(q_2)A_\nu^c(-q_1-q_2)
 \nnnl
&+\hat{Z}_1\,g\, T^a_{ij}(2q_{2\mu}+q_{1\mu})A_\mu^a(q_1)\bar{\varphi}^i(q_2)\varphi^j(-q_1-q_2)\Bigg)+\ldots\ ,
\end{align}
where $T^a_{ij}$ are the Hermitian generators in the fundamental representation of $SU(N_c)$ with the structure constants $f^{abc}$.
The dots correspond to four-point vertices and terms with Faddeev-Popov ghosts. The former are dropped in our truncation and the latter do not appear in the DSEs considered here, which are those of the scalar two-point function and of the scalar-gauge field vertex. The full DSE of the two-point function can be found, for example, in ref. \cite{Fister:2010yw}. Here we neglect for both DSEs all diagrams containing four-point functions which renders the scalar gap equation diagrammatically equal to the quark gap equation. This truncation is also motivated by the fact that two-loop diagrams are subleading in the UV.

\begin{figure}[tb]
\includegraphics[width=\textwidth]{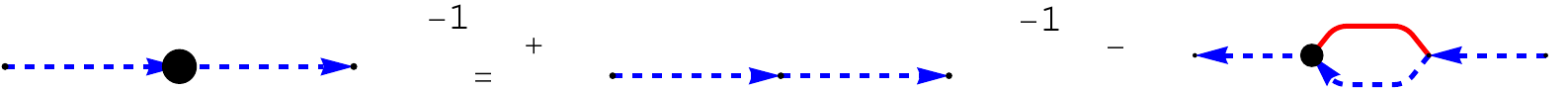}
\vskip3mm
\includegraphics[width=\textwidth,clip]{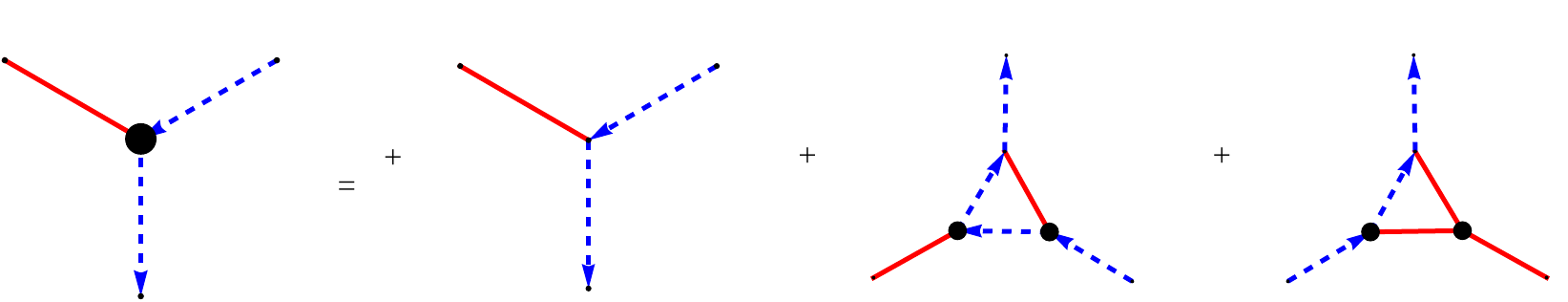}
\begin{center}
\caption{\label{fig:scalarDSEs}\textit{Top:} The truncated DSE of the scalar two-point function. The full DSE can be found, for example, in ref. \cite{Fister:2010yw}. \textit{Bottom:} The truncated DSE of the scalar-gauge field vertex. The second and third diagrams of the vertex DSE are called Abelian and non-Abelian diagrams, respectively.
Gauge fields have red, continuous lines and scalar fields blue, dashed lines. Thick blobs denote dressed vertices. All internal lines are dressed propagators.}
\end{center}
\end{figure}

In the following we will focus on the scalar sector of the theory, viz. the scalar propagator and the scalar-gauge field vertex. As can be seen from the  truncated DSEs of the scalar two-point function and the scalar-gauge field vertex in \fref{fig:scalarDSEs} we have the following four quantities left in our truncation: the propagators of the scalar and the gauge fields, the three-gauge field vertex and the scalar-gauge field vertex. 

The scalar propagator and the scalar-gauge field vertex are parametrized as
\begin{eqnarray}\label{eq:scalgreen}
 \left(D_s\right)_{mn}(p) & = & \frac{A_s(p^2)}{p^2}\delta_{mn}\ ,\\
 \Gamma^{\bar{s}sA,a}_{mn,\mu}(p_1, p_2) & = & g\,T^a_{mn}\left(A_{p_1}(p_1^2,p_2^2,z)\, p_{1\mu} + A_{p_2}(p_1^2,p_2^2,z)\,p_{2\mu}\right)\ ,\nonumber
\end{eqnarray}
where in the case of the vertex $p_1$ and $p_2$ are the momenta of the scalar particles and $z=p_1\cdot p_2/(|p_1| |p_2|)$. Introducing a sharp momentum cutoff $\Lambda$ to regularize the
self-energy contributions we need to approximate the three dressing functions

\begin{eqnarray}\label{eq:scaldress}
 A_s: \left[0,\Lambda^2\right] & \rightarrow & \mathbb{R}\ ,\\
 A_{p_1,p_2}: \left[0,\left(\Lambda/2\right)^2\right]^2\times\left[-1,1\right] & \rightarrow & \mathbb{R}\ ,
\end{eqnarray}
which will be done by linear interpolation for $A_{p_1,p_2}$ and linear interpolation as well as Chebyshev expansion in the case of $A_s$. Choosing the cutoff in the vertex to be smaller by a factor of two has the effect that only the dressing functions $A_{p_1,p_2}$ will be called at large momenta outside their domain when evaluating the self-energy integrals. We will approximate them with their bare value $A_{p_1,p_2} = \hat{Z}_1$ where necessary.

Additional information is required for the gauge field propagator and the three-gauge field vertex. For the latter we use for simplicity
\begin{align}\label{eq:AAA}
 \Gamma^{AAA,abc}_{\mu\nu\rho}(p_1, p_2, p_3) & = \frac{Z_3}{\tilde{Z}_3}\Gamma^{AAA,abc,(0)}_{\mu\nu\rho}(p_1, p_2, p_3) ,
\end{align}
where finiteness of the ghost-gauge field vertex in Landau gauge $\tilde{Z}_1=1$ and the Slavnov-Taylor identity $Z_1/\tilde{Z}_1=Z_3/\tilde{Z}_3$ \cite{Marciano:1977su} have been used. The bare vertex $\Gamma^{AAA,abc,(0)}_{\mu\nu\rho}(p_1, p_2, p_3)$ is given in \eref{eq:bareAAA}. For the dressing function $Z(p^2)$ of the Landau gauge field propagator we employ a fit to the solution of the ghost-gluon system obtained within the DSE framework  provided in ref. \cite{Fischer:2003rp} (see also section \ref{sec:YM}):
\begin{eqnarray}\label{eq:Z}
  Z(x) & = & \left(\frac{\alpha(x)}{\alpha(\mu^2)}\right)^{-\gamma}R^2(x)\ ,\\
  R(x) & = & \frac{c\left(\frac{x}{\Lambda_{QCD}^2}\right)^\kappa+d\left(\frac{x}{\Lambda_{QCD}^2}\right)^{2\kappa}}{1+c\left(\frac{x}{\Lambda_{QCD}^2}\right)^\kappa+d\left(\frac{x}{\Lambda_{QCD}^2}\right)^{2\kappa}}\ ,\nonumber
\end{eqnarray}
where $c=1.269$, $d=2.105$, $\gamma=-13/22$, $\kappa=0.5953$, $\Lambda_{QCD}=0.714$ GeV and

\begin{eqnarray}\label{eq:alpha}
 \alpha(x) & = & \frac{\alpha(0)}{\text{ln}\left[e+a_1\left(\frac{x}{\Lambda_{QCD}^2}\right)^{a_2}+b_1\left(\frac{x}{\Lambda_{QCD}^2}\right)^{b_2}\right]}\ ,
\end{eqnarray}
with $a_1=1.106$, $a_2=2.324$, $b_1=0.004$, $b_2=3.169$ and $\alpha(0)=8.915/N_c$, $N_c$ being the number of colors which we take to be three. Note that the renormalization constants $Z_3,\tilde{Z}_3$ can be calculated from the running coupling $\alpha$ as described in \cite{Fischer:2003rp}.

\begin{figure}[tb]
\includegraphics[width=\textwidth]{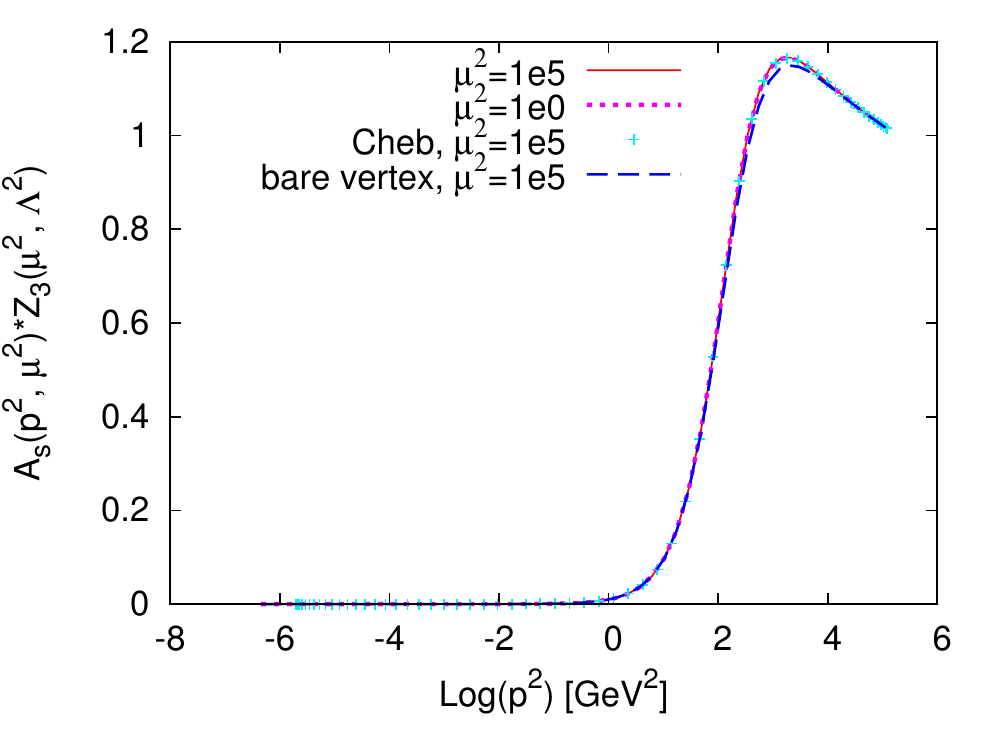}
\begin{center}
\caption{\label{fig:scalarProp}Dressing function of the propagator times its renormalization constant. Multiplicative renormalizability of the system propagator and vertex is clearly visible as the results obtained with two different choices of $\mu^2$ are indistinguishable. Independence from the approximation method is seen from the results obtained with the Chebyshev expansion method. For comparison the result with a bare scalar-gauge field vertex is included.}
\end{center}
\end{figure}

For the present system the iteration procedure is very stable and we can start from a massless bare scalar propagator $A_s\equiv 1$ and a bare scalar-gauge field vertex. As previously mentioned the integrals are regularized via an ultraviolet cutoff. To determine the renormalization constants of the scalar propagator we fix the values $A_s(\mu^2)$ and $Z_m m^2$. Furthermore the vertex will be renormalized by enforcing the Slavnov-Taylor identity $\hat{Z}_1/\hat{Z}_3 = \tilde{Z}_3/\tilde{Z}_3$. With this prescription the system is then multiplicatively renormalizable, i.~e., the expressions $\hat{Z}_3A_s$ and $(\hat{Z}_1)^{-1}A_{p_1,p_2}$ are independent of the
renormalization point $\mu^2$. Results confirming this for the propagator are shown in fig. \ref{fig:scalarProp}.

We can here also illustrate how extending truncations complicates the system of equations: A simple truncation takes into account only the scalar propagator. In this case we need the gauge field propagator and the scalar-gauge field vertex as input and we only have to calculate one integral. Its integrand is comparatively simple. Going only one step further by including also the scalar-gauge field vertex requires solving in total for three dressing functions (the vertex has two and the propagator one) by calculating five integrals, whereby the complexity of the integrands has also increased considerably.

We should mention that the employed truncation is not state of the art but it is sufficient to illustrate many features of \textit{CrasyDSE}. More elaborate results for Yang-Mills theory coupled to a fundamental scalar will be presented elsewhere \cite{Alkofer:2011up}.

\subsection{Generating the \textit{C++} files with the integrands}

We turn now to the creation of the kernels with \textit{Mathematica}. The subsequent explanations follow the notebook \textit{CrasyDSE\_YM+scalar.nb}.
To initialize the required functions and the expressions for the DSEs we evaluate the initialization cells with
\begin{verbatim}
FrontEndTokenExecute["EvaluateInitialization"]
\end{verbatim}
or via the menu entry \textit{Evaluation $\rightarrow$ Evaluate Initialization cells}. Now the variables containing the expressions of the integrals are defined:
\tw{gapAlgProjLoop\-Integrand} is the integral of the gap equation and \tw{vertex\-Abelian\-Projp1\-Final}, \tw{vertex\-NonAbelian\-Projp1Final}, \tw{vertex\-Abelian\-Projp2\-Final} and \tw{vertex\-Non\-Abe\-lian\-Projp2\-Final} correspond to the integrals of the Abelian and non-Abelian diagrams projected onto the external momenta $p_1$ and $p_2$. Furthermore the package \textit{CrasyDSE} is loaded.

Basically for the generation of the \textit{C++} files only one function is needed. However, before we can use it we need to define several expressions. They are split into lists containing parameters, momentum variables and dressing function names.
In the following the order of the elements in lists is very important. Thus one has to be very careful when one changes something later on, because this could lead to inconsistencies with the \textit{C++} code. 

In the gap equation two parameters appear: the number of colors $N_c$ and the coupling constant $g$. We put them both into a list:
\begin{verbatim}
parasGap = {Nc, g};
\end{verbatim}
Furthermore we specify the external (\tw{ps}$:=p^2$) and internal (\tw{qs}$:=q^2$, \tw{ct}$:=\cos(\varphi)=p\cdot q/|p||q|$) variable names:
\begin{verbatim}
extVarsGap = {ps};
intVarsGap = {qs, ct};
\end{verbatim} 
Although we did not introduce abbreviations of momentum combinations for the gap equation, we need to define a variable for this, because we will need it later:
\begin{verbatim}
extraVarsCListGap = {};
\end{verbatim} 
Finally we have to specify which dressings appear in each equation. They are separated into two lists, one for the dressings of the Green function we are calculating and one for all other dressings belonging to other Green functions. The propagator has only one dressing function $D_s$, so the first list contains one item only:
\begin{verbatim}
dressingsGap = {Ds};
\end{verbatim}
But the gap equation also depends on other dressing functions, namely on the one of the gauge field, $D_A$, and on the two of the scalar-gauge field vertex, $D_{As\bar{s}}^{(1)}$ and $D_{As\bar{s}}^{(1)}$. All dressings belonging to the same Green function have to be grouped together in one sublist:
\begin{verbatim}
otherGreenFuncsGap = {{DA}, {DAssb1, DAssb2}};
\end{verbatim}
These are the lists required for the gap equation as input for generating the \textit{C++} files.

The lists for the vertex equation have the same structure and we only list them here:
\begin{verbatim}
parasVertex = {Nc, g};
extVarsVertex = {p1s, p2s, ca};
intVarsVertex = {qs, ct1, ct2};
extraVarsCListVertex = {{p1p2, ca Sqrt[p1s p2s]},
  {p1q, Sqrt[p1s qs] Cos[ct2]},
  {p2q, (ca ct2 + Sqrt[1 - ca^2] ct1 Sqrt[1 - ct2^2]) Sqrt[p2s qs]},
  {p1mqs, p1s - 2 p1q + qs},
  {p2mqs, p2s - 2 p2q + qs},
  {p1mp2s, p1s + p2s - 2 Sqrt[p1s p2s] ca}};
dressingsVertex = {DAssb1, DAssb2};
otherGreenFuncsVertex = {{DA}, {Ds}, {DAAA}};
\end{verbatim} 
Note that the vertex has two dressings by itself and depends on three other Green functions. Furthermore we provided with the list \verb|extraVarsCListVertex| the definitions of employed abbreviations, e.~g., \tw{p1p2}$:=p_1\cdot p_2=\cos(\alpha)|p_1||p_2|$.

Before we generate the \textit{C++} files we split off numeric coefficients of the integrands. We call the resulting expressions \textit{kernels} and \textit{coefficients}. Instead of doing this by hand, we use the function \tw{splitIntegrand}. It takes as arguments an expression and a list of variables. Everything in the overall factor that does not contain a variable will be put into the coefficient:
\begin{verbatim}
{coeffGap, kernelGap} = 
 splitIntegrand[gapAlgProjLoopIntegrand, 
    Join[extVarsGap, extraVarsCListGap[[All, 1]], intVarsGap]] /. 
   Z1h :> 1 // Simplify

--> {(g^2 (-1 + Nc^2))/(
 8 Nc \[Pi]^3), (1/pplusqs)(1 - ct^2)^(3/2)
   DA[qs] (DAssb1[ps, 
    pplusqs, -((ps + ct Sqrt[ps qs])/Sqrt[pplusqs ps])] - 
   DAssb2[ps, 
    pplusqs, -((ps + ct Sqrt[ps qs])/Sqrt[pplusqs ps])]) Ds[pplusqs]}
\end{verbatim}
We discarded the renormalization function Z1h here since its implementation is handled manually.

For the vertex we do the same but bear in mind the following structure: Both for coefficients and kernels every loop integral is treated as a single expression, and for every equation all loops are grouped into lists. The syntax of the list of coefficients or kernels is thus
\begin{verbatim}
{{loop 1 of eq. 1, loop 2 of eq. 1, ...},
 {loop 1 of eq. 2, loop 2 of eq. 2, ...}, ...}
\end{verbatim} 
For the first and second projections we split the integrands as follows:
\begin{verbatim}
{coeffsVertexProjp1, kernelsVertexProjp1} = Transpose[
 splitIntegrand[#, 
  Join[extVarsVertex, extraVarsCListVertex[[All, 1]], 
    intVarsVertex]] & /@ {vertexAbelianProjp1Final, 
     vertexNonAbelianProjp1Final} /. {Z1h :> 1, Z1 :> 1} // Simplify];
{coeffsVertexProjp2, kernelsVertexProjp2} = 
 Transpose[
  splitIntegrand[#, 
   Join[extVarsVertex, extraVarsCListVertex[[All, 1]], 
    intVarsVertex]] & /@ {vertexAbelianProjp2Final, 
     vertexNonAbelianProjp2Final} /. {Z1h :> 1, Z1 :> 1} // Simplify];
\end{verbatim} 
Again we have discarded the renormalization functions.
The final lists of kernels and coefficients are
\begin{verbatim}
kernelsVertex = {kernelsVertexProjp1, kernelsVertexProjp2};
coeffsVertex = {coeffsVertexProjp1, coeffsVertexProjp2}

--> {{g/(16 Nc Pi^3), -((g Nc)/(32 Pi^3))}, 
 {g/(16 Nc Pi^3), -((g Nc)/(32 Pi^3))}}
\end{verbatim} 
We show the coefficients explicitly. One can easily spot the $1/N_c$ and $N_c$ dependences of the Abelian and non-Abelian diagrams, respectively.

Finally we have everything to generate the \textit{C++} code. We do so with the function \tw{exportKernels}:
\begin{verbatim}
exportKernels[{FileNameJoin[{NotebookDirectory[], ".."}],
  "kernelsAll"},
 {"scalar_QCD.hpp"},
 {{{"coeffGap", "kernelGap"},
   {coeffGap},
   {kernelGap},
   dressingsGap,
   otherGreenFuncsGap,
   parasGap,
   extVarsGap,
   intVarsGap,
   extraVarsCListGap},
  {{"coeffsVertex", "kernelsVertex"},
   coeffsVertex,
   kernelsVertex,
   dressingsVertex,
   otherGreenFuncsVertex,
   parasVertex,
   extVarsVertex,
   intVarsVertex,
   extraVarsCListVertex}
  }]
\end{verbatim}
It will create two files \textit{kernelsAll.hpp} and \textit{kernelsAll.cpp}. The filenames are determined by the first argument where we also indicate that the files should be exported to the parent directory.
The second argument here is the name of an additional header file which contains functions specific to this example.
The third argument contains all the information we gathered above: It is a list where every item corresponds to one DSE. For every DSE we have the following entries:
\begin{itemize}
 \item The \textit{C++} names of the functions containing the coefficients and the kernels.
 \item A list with the expression(s) for the coefficient(s).
 \item A list with the expression(s) for the kernel(s).
 \item The list of dressings for this Green function.
 \item The list of dressings from other Green functions.
 \item The list of parameters.
 \item The list of external variables.
 \item The list of internal variables.
 \item The list of extra variables.
\end{itemize}
Note that without specifying a path in the first argument of \texttt{exportKernels} the files will be created in the directory of the notebook.

We want to mention here the function \tw{functionToString}, which is used by \tw{exportKernels} but can also be used directly by the user. It creates a string of the expression given as its argument similar to the \textit{Mathematica} function \tw{CForm}, but it replaces some common functions like \tw{Power} or \tw{Sin} by its \textit{C++} counterparts \tw{pow} or \tw{sin}. If a function is not included, it can be added by hand, for example:
\begin{verbatim}
functionToString[a^b + Sin[b]/10 - 5 Sinh[a], {Sinh :> sinh}]

--> 0.1*(sin(b)) + -5.*(sinh(a)) + (pow(a, b))
\end{verbatim}

This finishes our work in \textit{Mathematica} and we proceed with the \textit{C++} code.

\subsection{Numerical code}

The \textit{C++} code, contained in \textit{scalar\_main.cpp}, is extensively commented and every variable that appears is described directly in the file. Here we only give a rough overview of the required initializations.

In the file \textit{scalar\_main.cpp} first the model and then all Green functions are initialized. For the former the definitions are as simple as providing numeric values for some parameters, e.~g., $N_c=3$. All Green functions are defined as a \texttt{dse} structure, which contains a pointer to \texttt{mod} which is reserved for hosting model parameters, see also \fref{fig:structures}. Since the gauge field propagator and the three-gauge field vertex are given by ans\"atze the only additional information these structures need are the corresponding definitions. The dynamically calculated Green functions also contain an array of \texttt{quad} structures, namely one for each different integration. Consequently variables like the numbers of integration points and the quadrature types have to be initialized. We also have to provide starting expressions for the dressings and information on the other Green functions contained in a DSE - handled via the array \texttt{otherGF}. For example, for the gap equation these are the gauge field propagator and the scalar-gauge field vertex. In the \textit{C++} code dressing functions do not have a specific name, but are just collected in the function \texttt{dress} where it is important at all times to maintain the same assignment of the dressing functions as in the notebook. The integrands of the self-energy are defined in the kernels file created with \textit{CrasyDSE\_YM+scalar.nb}. Also a renormalization procedure has to be defined. Finally, the iteration is done with the function \texttt{meta\_solve\_iter}.

\section{Landau gauge Yang-Mills theory}
\label{sec:YM}

As a second example we use pure Yang-Mills theory which requires some different methods. The most obvious change is that we use a Newton's method instead of a direct fixed point iteration. Again we provide the complete \textit{Mathematica} and \textit{C++} code together with the program. However, as the creation of the kernel files is rather similar to the case of the previous section we refrain from showing any details.

\subsection{Truncation and ans\"atze}

As in the last section we will employ the Landau gauge. The DSE system truncated at the level of propagators has been investigated with DSEs for some time now, see, for example, \cite{vonSmekal:1997vx,vonSmekal:1997is,Atkinson:1997tu,Fischer:2002hn,Aguilar:2008xm,Fischer:2008uz}. We will here reproduce the solutions of refs. \cite{Fischer:2002hn,Fischer:2003zc, Fischer:2008uz}. Besides employing a Newton procedure to solve this set of equations another difference to the previous section lies in the renormalization procedure: Here we work with subtracted DSEs.

\begin{figure}[tb]
\includegraphics[scale=0.9]{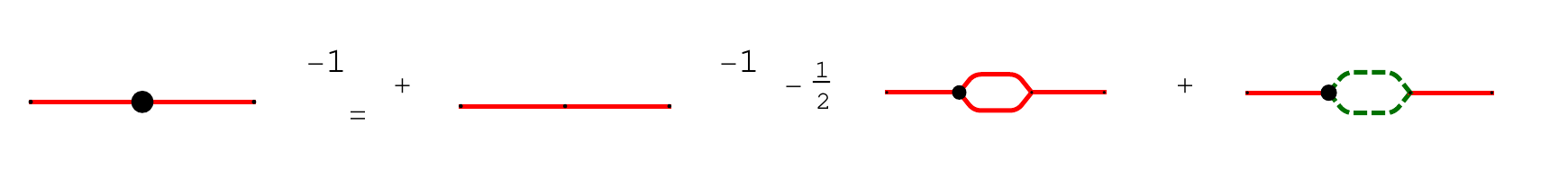}
\vskip3mm
\includegraphics[scale=0.75]{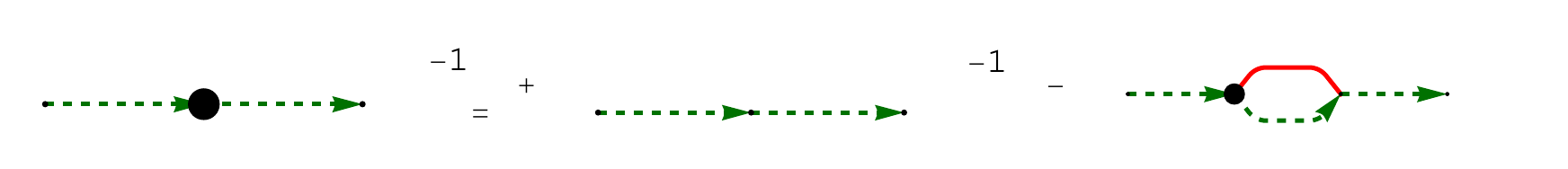}
\begin{center}
\caption{\label{fig:DSEs_YM}The truncated two-point DSEs of pure Yang-Mills theory. Gluons have red, continuous lines and ghost fields green, dashed lines. Thick blobs denote dressed vertices. All internal lines are dressed propagators.}
\end{center}
\end{figure}

The system we investigate consists of the ghost and gluon two-point DSEs. The former is used without change, while the gluon DSE is truncated \cite{vonSmekal:1997vx,vonSmekal:1997is}: We neglect all diagrams involving a bare four-gluon vertex, i.e., the tadpole diagram and all two-loop diagrams. They are subleading in the UV and it was shown analytically for the scaling solution that they are also subleading in the IR \cite{Fischer:2009tn}. The remaining unknown quantities are the ghost-gluon and the three-gluon vertices for which we use suitable ans\"atze. The truncated set of DSEs is depicted in \fref{fig:DSEs_YM}.

\begin{figure}[tb]
\begin{center}
 \includegraphics[width=\linewidth]{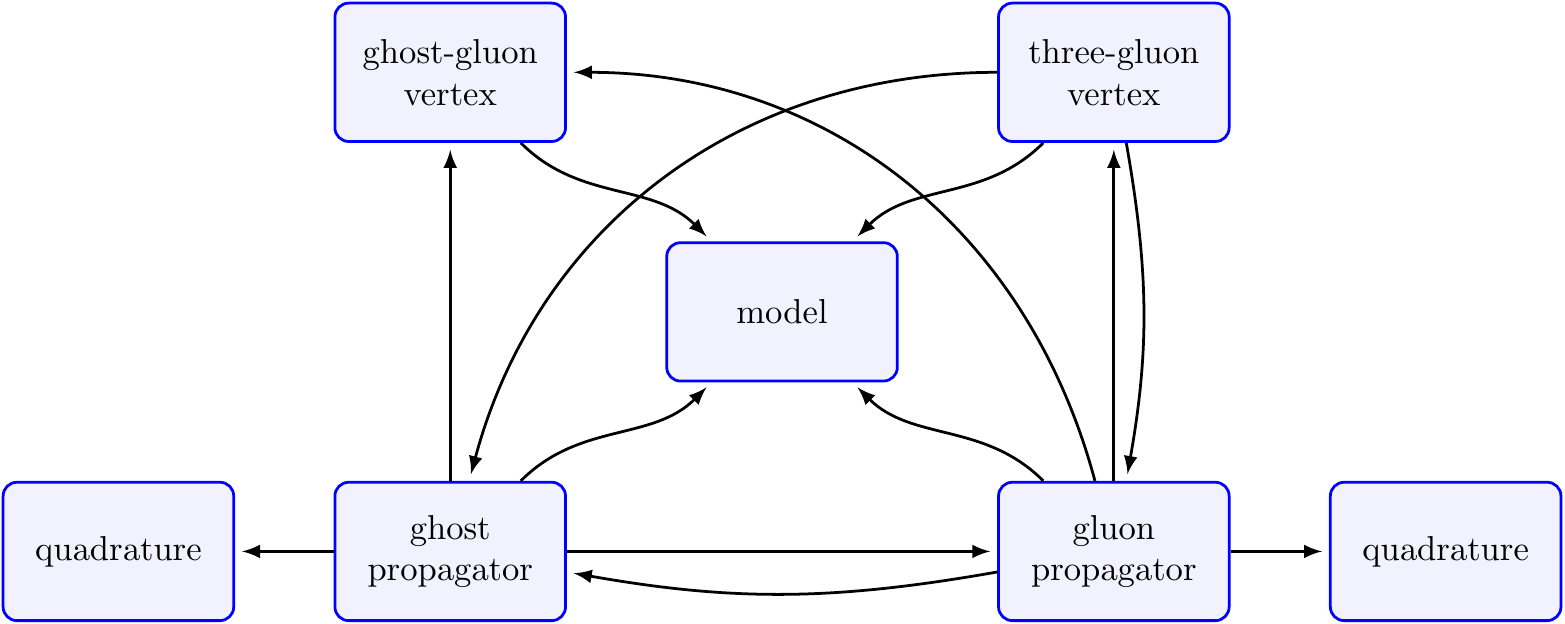}
\caption{\label{fig:structures_YM}Schematic illustration of structures defined in the code and their dependencies for the pure Yang-Mills system.}
\end{center}
\end{figure}

The ghost and gluon propagators are given by
\begin{align}
 D^{ab}_{\mu\nu}(p)&=\de^{ab}\left( g_{\mu\nu}-\frac{p_\mu p_\nu}{p^2} \right) \frac{Z(p^2)}{p^2},\\
 G^{ab}(p) & = -\de^{ab}\frac{G(p^2)}{p^2}.
\end{align}
For the ghost-gluon vertex $\Gamma^{A\bar{c}c,abc}_{\mu}(p_1, p_2, p_3)$ the bare version,
\begin{align}
\Gamma^{A\bar{c}c,abc,(0)}_{\mu}(p_1, p_2, p_3) = i\,g\,f^{abc} p_{2\mu},
\end{align}
is used as motivated originally by an argument of Taylor. However, several studies both on the lattice \cite{Cucchieri:2008qm} and in the continuum \cite{Lerche:2002ep,Schleifenbaum:2004id} confirmed this to be a very reliable ansatz. For the full three-gluon vertex $\Gamma^{AAA,abc}_{\mu\nu\rho}(p_1, p_2, p_3)$ we use the tensor structure of the bare vertex $\Gamma^{AAA,abc,(0)}_{\mu\nu\rho}(p_1, p_2, p_3)$ amended by a dressing $D^{AAA}(p_1^2, p_2^2, p_3^2) $ that guarantees the correct UV behavior of the gluon dressing function \cite{Fischer:2002hn}:
\begin{align}\label{eq:bareAAA}
 \Gamma^{AAA,abc,(0)}_{\mu\nu\rho}(p_1, p_2, p_3) & = i\,g\,f^{abc} \left(g_{\mu\nu}(p_2-p1)_\rho+g_{\nu\rho}(p_3-p2)_\mu+g_{\rho\mu}(p_1-p_3)_\nu \right) ,\\
 \Gamma^{AAA,abc}_{\mu\nu\rho}(p_1, p_2, p_3) & = \Gamma^{AAA,abc,(0)}_{\mu\nu\rho}(p_1, p_2, p_3) D^{AAA}(p_1^2, p_2^2, p_3^2),\\
 D^{AAA}(p_1^2, p_2^2, p_3^2) & = \frac{1}{Z_1}\frac{\left(G(p_2^2) G(p_3^2) \right)^{1-a/\de-2a}}{\left(Z(p_2^2)Z(p_3^2)\right)^{1+a}}.
\end{align}
$a$ is a parameter chosen as $3\delta$, where $\de$ is the anomalous dimension of the ghost propagator. $Z_1$ is the renormalization constant of the three-gluon vertex. The dependencies of all Green functions on each other 
are shown in \fref{fig:structures_YM}.

An important issue of the gluon DSE are spurious quadratic divergences. They appear because we employ a numerical cutoff as UV regularization which breaks gauge invariance. There are several ways to deal with them, see, for example, \cite{Fischer:2002hn,Fischer:2008uz,Pennington:2011xs}. Here we subtract an additional term in the kernel of the gluon loop in the gluon DSE  \cite{Fischer:2002hn}. The derivation of the DSEs with \textit{DoFun} and the projection to scalar quantities are described in the notebook \textit{DoFun\_YM\_4d.nb} and the creation of the kernel files in \textit{CrasyDSE\_YM\_4d.nb}. For details we refer to them.

\subsection{Renormalization and solution}
\label{sec:YM_ren_sol}

For the present system we will use subtracted DSEs, i.~e., we subtract from a DSE at external momentum $p$ the DSE at a fixed external momentum $p_0$:
\begin{align}
 D^{-1}(p^2) = Z^{-1}+\Pi(p^2) \quad \Rightarrow \quad D^{-1}(p^2) =  D^{-1}(p_0^2)+\Pi(p^2) -\Pi(p_0^2).
\end{align}
Thus we can trade the renormalization constant $Z$ for specifying the value of the dressing function at the subtraction point $p_0$. For the gluon propagator we choose the subtraction point $p_0$ at sufficiently high momenta since we expect the two-point function to be divergent at low momenta. For the ghost, however, it is most advantageous to specify the dressing at zero momentum. These conditions are boundary conditions for the integral equations. As it turns out two different types of solutions can be found depending on the value of the ghost dressing at zero momentum: Choosing finite values a solution of the family of decoupling solutions emerges \cite{Boucaud:2008ji,Aguilar:2008xm,Fischer:2008uz}, while with an infinite zero-momentum dressing we get the scaling solution \cite{vonSmekal:1997vx,Fischer:2008uz}. The former has a finite gluon propagator and a finite ghost dressing function at zero momentum and the latter an IR vanishing gluon propagator and an IR divergent ghost dressing function. Thereby the divergence of the ghost dressing and the vanishing of the gluon dressing can be described by power laws whose exponents $\de_{gh}$ and $\de_{gl}$, respectively, are related by $\de_{gl}+2\de_{gh}=0$, whereby $\de_{gh}:=\ka=0.595353$ can be calculated analytically \cite{Zwanziger:2001kw,Lerche:2002ep}.

There are several choices at which point of the calculation the subtraction of a DSE can be performed. For illustration purposes we employ a different one for each propagator: For the ghost we use the subtracted expression in the kernel file. This is advantageous because the limit of vanishing external momentum can be done analytically but is problematic numerically. For the gluon propagator, on the other hand, we only create the unsubtracted expressions. The subtraction is then performed with a function in \textit{C++}. Here the subtraction is numerically unproblematic.

The specific renormalization procedure has to be taken into account in the renormalization function in the \textit{C++} code. Furthermore, it is important to note that this function does not calculate the right-hand side of a DSE as in the example of the scalar system but the difference between the right- and left-hand side of a (subtracted) DSE:
\begin{align}
 E(p^2):=-D^{-1}(p^2) +  D^{-1}(p_0^2)+\Pi(p^2) -\Pi(p_0^2).
\end{align}
The reason is the employed Newton procedure as described in section \ref{sec:Newton} which attempts to bring $E(p^2)$ to zero. $E(p^2)$ is calculated for every external momentum and saved as an array of the DSE structure.

The behavior of the dressing functions in the IR and UV is known analytically. In the intermediate regime, between two given momenta $\epsilon$ and $\Lambda$, above and below the IR and UV cutoffs, respectively, they are expressed by an expansion in $N$ Chebyshev polynomials\footnote{The exponential is chosen due to better convergence properties. For such an expansion see also, for example, ref. \cite{Bloch:1995dd,Maas:2005xh}.}:
\begin{align} 
 G_{IM}(p^2)&=\exp{\sum_{i=0}^{N-1}c^{(gh)}_i \, T_i(M(p^2))},\\
 Z_{IM}(p^2)&=\exp{\sum_{i=0}^{N-1}c^{(gl)}_i \, T_i(M(p^2))},
\end{align}
where $M(p^2)$ maps the regime $[\epsilon, \Lambda]$ to $[-1,1]$.
For momenta below $\epsilon$ we employ a power law with the given exponent and the coefficient calculated from the lowest known point in the Chebyshev expansion:
\begin{align}
 G_{IR}(p^2) &= A^{(gh)}\,(p^2)^{\de_{gh}},\\
 Z_{IR}(p^2) &= A^{(gl)}\,(p^2)^{\de_{gl}}.
\end{align}
For momenta higher than $\Lambda$ an extrapolation in agreement with the UV behavior is chosen:
\begin{align}
 G_{UV}(p^2)&=G(s^2)(w\log(p^2/s^2)+1)^\delta,\\
 Z_{UV}(p^2)&=Z(s^2)(w\log(p^2/s^2)+1)^\gamma.
\end{align}
$s$ is the highest momentum at which the Chebyshev expansion is known, $\de$ or $\gamma$  are the anomalous dimensions of the ghost and gluon, respectively, and $w=11\,N_c\,\alpha(s)\, G(s)^2\, Z(s)/12 \pi$. $\alpha(p^2)$ is a possible non-perturbative definition of the running coupling \cite{Alkofer:2002ne,vonSmekal:2009ae}:
\begin{align}
 \alpha(p^2):=\alpha(\mu^2) G(p^2)^2 Z(p^2).
\end{align}
The value of $\alpha(\mu)$ is an input parameter and sets the scale: At $\mu$ we have $G(\mu^2)^2Z(\mu^2)=1$.

As starting functions for the propagator dressings in the intermediate regime we use
\begin{align}
 Z_{ans}(p^2)&= f_{IR}(p^2)^2 (p^2)^{\de_{gl}}+c_{UV}\cdot f_{UV}(p^2),\\
 G_{ans}(p^2)&= f_{IR}(p^2) (p^2)^{\de_{gl}}+1,
\end{align}
with the IR and UV damping factors given by
\begin{align}
 f_{IR}(p^2)& = \frac{L_{IR}}{L_{IR}+p^2},\\
 f_{UV}(p^2)& = \left(\frac{p^2}{L_{UV}+p^2}\right)^2,
\end{align}
where $L_{IR}$ and $L_{UV}$ are dimensionful parameters conveniently set to $1$.
The parameter $c_{UV}$ can be used to adjust the starting function in order to speed up convergence. Here it is chosen as $1$.
These ans\"atze respect the qualitative IR behavior which leads to a faster convergence than 
starting with constant functions. In general the Newton procedure becomes more stable when using starting functions close to the solution. 
The starting functions being not differentiable at the UV matching poses no problem.

\begin{figure}
 \begin{center}
  \includegraphics[width=0.48\textwidth]{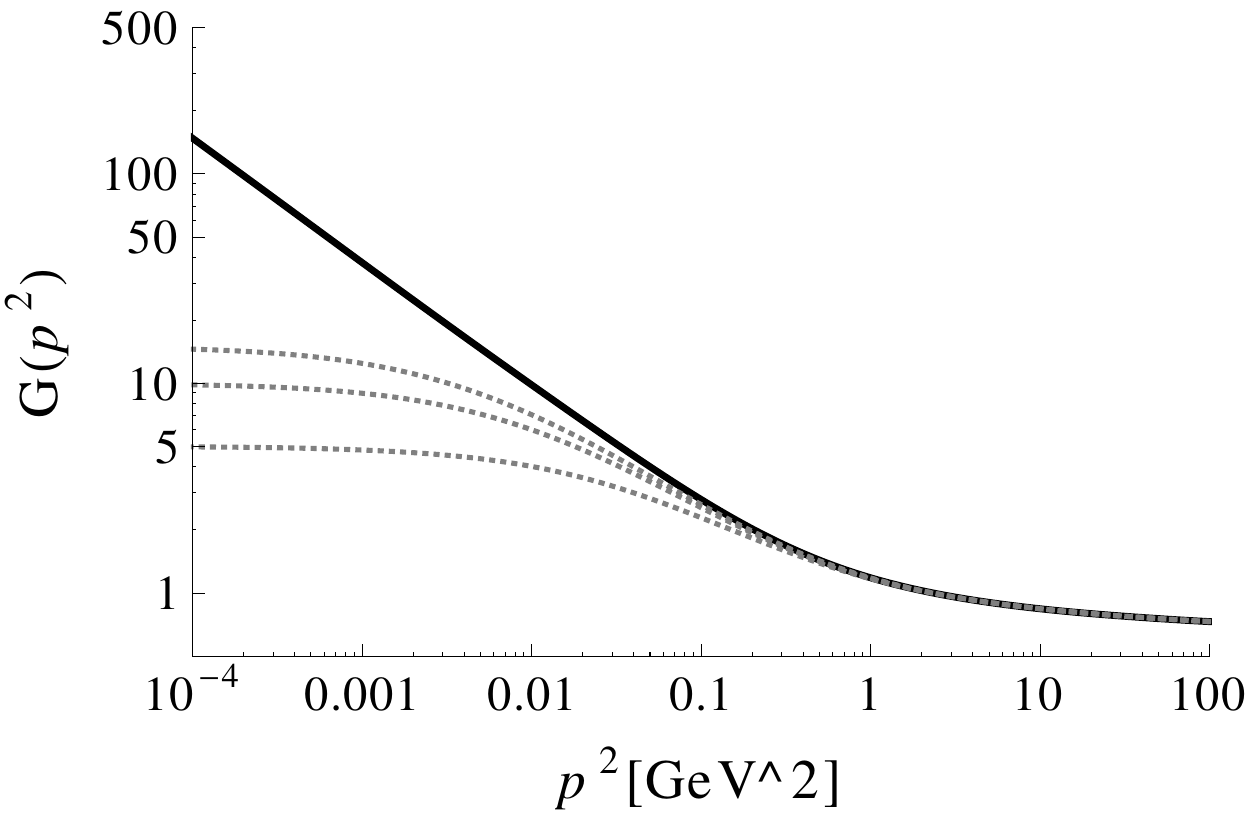}
  \includegraphics[width=0.48\textwidth]{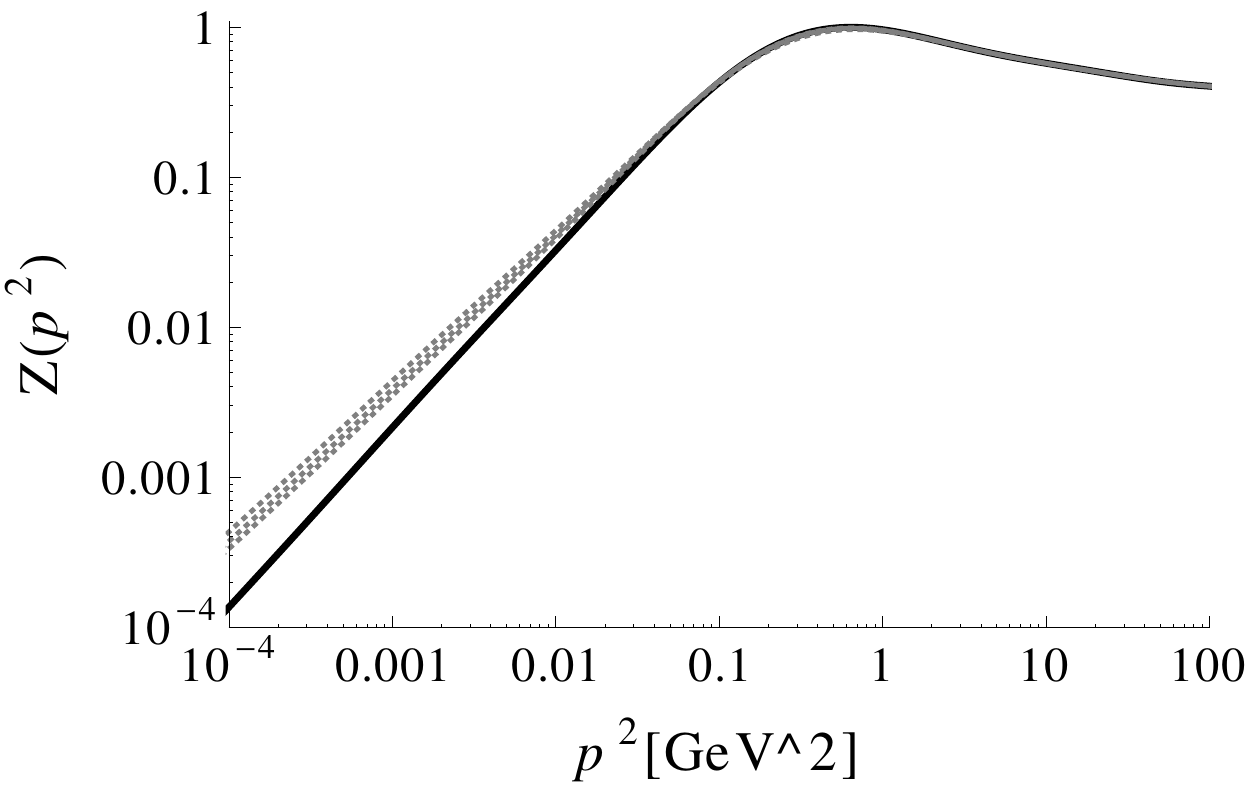}
 \end{center}
\caption{\label{fig:props_YM} Ghost (left) and gluon (right) dressing functions. The continuous line corresponds to the scaling solution, the dashed lines are solutions of the decoupling type.}
\end{figure}

For the integration we used Gauss-Legendre quadratures. Furthermore it was advantageous to split the radial integration at the value of the external momentum which requires setting \texttt{bound\_type} of all quadratures to \texttt{1}. Note that this has to be done after \texttt{init\_quad}, which sets the default value \texttt{0}. This allows a higher precision of the final result. However, 
this comes at the prize of slowing down the integration as the integration boundaries have to be calculated for every external momentum.

\begin{table}
\begin{center}
 \begin{tabular}{|l|c||l|c|}
  \hline
  parameter & value & parameter & value\\
  \hline
  \hline
  $N_c$ & 3 & $\kappa$ & $0.595353$\\
  \hline
  $\alpha(\mu^2)$ & $1$, $0.5$ & $h$ & $10^{-3}$\\
  \hline
  UV cutoff & $10^3$ &$\epsilon$ & $2\times 10^{-8}$\\
  \hline
  IR cutoff & $10^{-12}$ & $\Lambda$ & $0.99\times 10^3$\\
  \hline
  $\de$ & $-9/44$ & gluon subtraction point $p_0$ & $1.2$\\
  \hline
  $\gamma$ & $-13/44$ & value of gluon dressing at $p_0$ & $0.93$\\
  \hline
  $L_{IR}$ & $1$ & value of ghost dressing at $p=0$ & $0$, $5$, $10$, $25$\\
  \hline
  $L_{UV}$ & $1$ & & \\
  \hline
 \end{tabular}
\caption{\label{tab:YM_paras}Parameters for the calculation. Where several values are given see text for details.}
\end{center}
\end{table}

In \fref{fig:props_YM} we show the results of the calculations for different boundary conditions of the ghost. It is clearly visible that all  solutions coincide in the UV and only show their distinct behavior in the IR. In table \ref{tab:YM_paras} we provide the input parameters used for our calculations. The reached precision 
can be seen from how well the results fulfill the DSEs, i.~e., how close $E(p^2)$ approaches zero. Its norm goes with \texttt{double} precision down to about $10^{-6}$. Using \texttt{long double} variables instead this value can be made even lower.

\begin{figure}
 \begin{center}
  \includegraphics[width=0.48\textwidth]{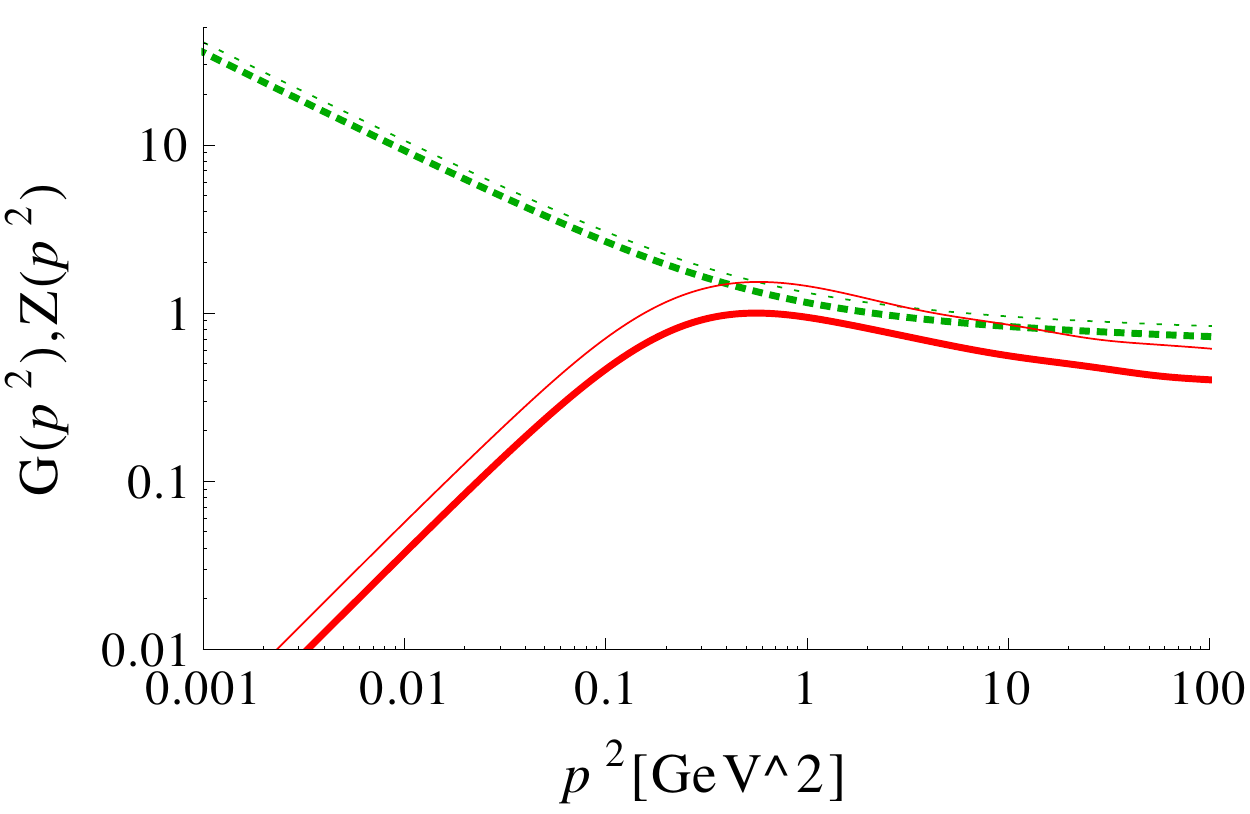}
  \includegraphics[width=0.48\textwidth]{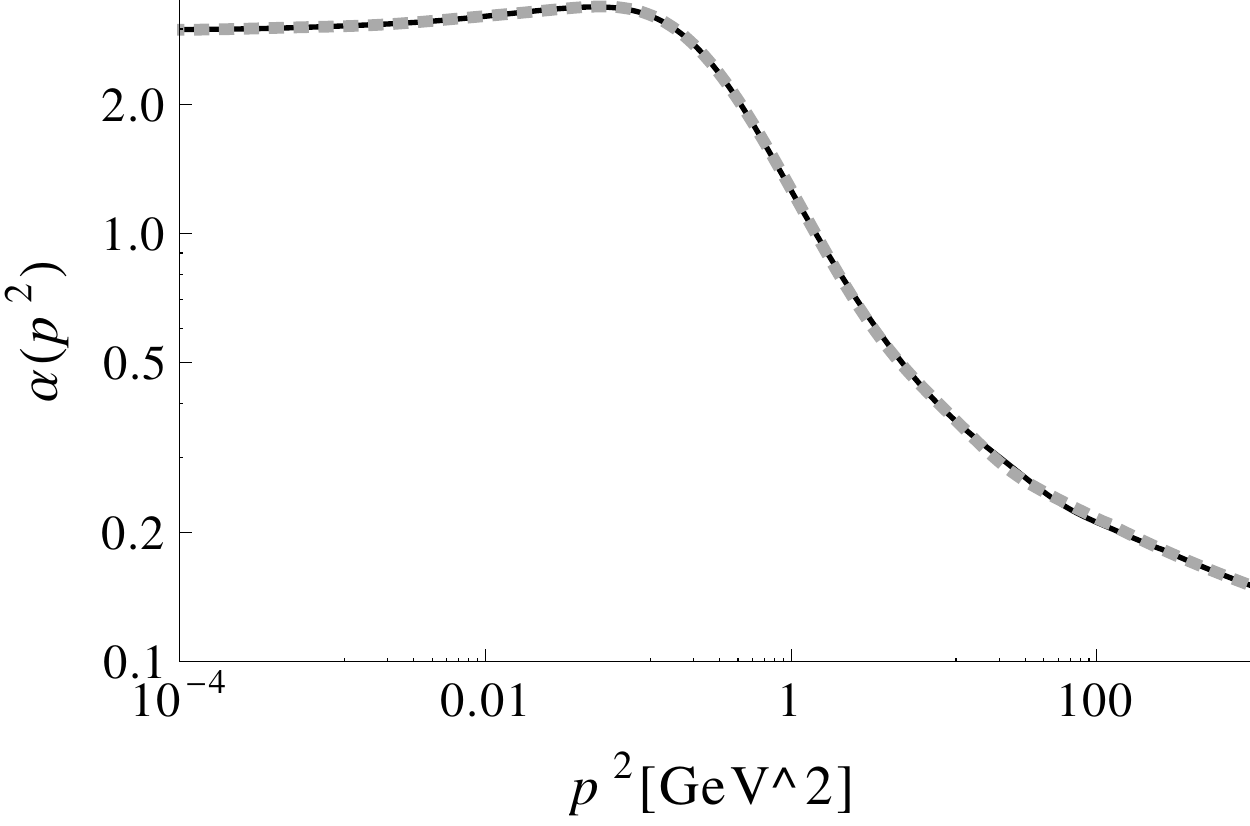}
 \end{center}
\caption{\label{fig:alphaComp} \textit{Left:} Ghost and gluon propagators. The continuous (red) lines corresponds to the gluon, the dashed (green) lines to the ghost. Thick and thin lines corresponds to different choices of $\alpha(\mu^2)$. \textit{Right}: Coupling for the two choices of $\alpha(\mu^2)$. The two lines (black straight and gray dashed) are almost indistinguishable.}
\end{figure}

We also tried a second choice for $\alpha(\mu^2)$ to test the code, namely $\alpha(\mu^2)=0.5$. As expected the propagators change by a constant factor due to multiplicative renormalizability, whereas the running coupling $\alpha(p^2)$ is independent of $\mu^2$, see, for example, \cite{Alkofer:2004it,Fischer:2008uz}. A comparison between $\alpha(\mu^2)=1$ and $0.5$ is depicted in \fref{fig:alphaComp}.

Finally we want to make some technical remarks: With these examples we only want to illustrate the basic use of \textit{CrasyDSE} so the code is not optimized and we expect that the runtime can be improved considerably. Another point is that we also tried a simple fixed point iteration but did not get a solution. This may indicate that this method is not suited for this problem.

\section{Summary and outlook}
\label{sec:outlook}

The strength of the framework provided by \textit{CrasyDSE} lies in its ability to handle large numbers of dressing functions. Thus one of its fields of application is the 
extension
of current truncation schemes by including higher vertices and/or enlarging the tensor bases of Green functions. For example, it is expected that going beyond the current truncation schemes in the Landau gauge makes DSE solutions more competitive to lattice solutions in the mid-momentum regime. Since the number of dressing functions increases at non-zero temperature and/or non-zero density corresponding calculations can profit from \textit{CrasyDSE} too. Finally there are also interesting cases for which no numerical calculations have been done successfully yet because their systems of DSEs are very complex. 

Combining \textit{CrasyDSE} with \textit{DoFun} there exists a sound framework for the treatment of all such systems of DSEs, from their derivations to their numeric solutions.
Furthermore, we plan to extend \textit{CrasyDSE} by adding further approximation methods or new solving algorithms to the ones given in table \ref{tab:methods} as required by individual cases.

\begin{table}[tb]
\begin{center}
 \begin{tabular}{|l|l|}
  \hline
  interpolations: & linear\\
  \hline
  expansions: & Chebyshev\\
  \hline\hline
  solution methods: & fixed point iteration, Newton\\
  \hline  \hline
  quadratures: & Gauss-Legendre, Chebyshev, Fejer, double exponential\\
  \hline
 \end{tabular}
\caption{\label{tab:methods}Currently implemented numerical methods.}
\end{center}
\end{table}

\section*{Acknowledgments}
We would like to thank Reinhard Alkofer, Christian S. Fischer, Markus Hopfer, Axel Maas and Lorenz von Smekal for valuable discussions. We are grateful to Axel Maas for a critical reading of an early version of this manuscript and to Christian S. Fischer for providing numerical results for comparison. Furthermore we want to thank Tina Katharina Herbst for a careful reading of the manuscript and useful comments.
MQH is supported by the Alexander von Humboldt foundation. MM acknowledges support by the Doktoratskolleg "Hadrons in Vacuum, Nuclei and Stars" of the Austrian science fund (FWF) under contract W1203-N16.

\appendix

\section{Initializing and accessing \texttt{x\_A} and \texttt{A}}
\label{sec:x_A-A}

Here we provide some details on accessing and initializing the arrays \texttt{double *x\_A} and \texttt{double *A} which are members of the structure \texttt{struct dse}. Details on their structure can be found as comments in \textit{DSE.cpp}. Here we only describe the functions to access them properly which should be sufficient for most applications. For every DSE we also have to define the number of external momenta and the number of dressing functions, \texttt{int dim\_x} and \texttt{int dim\_A}, respectively. For every external momentum the number of interpolation points or expansion coefficients has to be specified. The corresponding values form the array \texttt{int *n\_A}. As an introductory example for the use of these variables one can consider the interpolation example \textit{interp\_main.cpp}.

\subsection{The array \texttt{x\_A}}
\label{sec:x_A}

The array \texttt{x\_A} contains the interpolation points and/or Chebyshev nodes depending on the chosen interpolation method. 
In the case of Chebyshev interpolation only the discrete variables need to be initialized by hand in \texttt{x\_A}. 
However when using the DSE solving routines it is necessary that also the continuous Chebyshev nodes are stored in \texttt{x\_A}. These are automatically initialized in \texttt{x\_A} by calling the function \texttt{void Cheb\_init\_cont\_xA(void *dse\_param)}. In the provided examples \texttt{Cheb\_init\_cont\_xA} is called in the member \texttt{void (*init\_dress)} of the structure \texttt{dse} 
, e.~g., \texttt{scalprop\_init\_dress\_cheb} for the scalar propagator. Note that due to technical reasons the values for each continuous variable are ordered from low to high for a linear interpolation, but from high to low for a Chebyshev expansion. 

Independent of the interpolation method the discrete external momenta have to be initialized by the user in \texttt{x\_A}. In the case of linear interpolation this is also true for the continuous arguments\footnote{For Chebyshev interpolation this has to be done manually as well if one wants other external momenta than the Chebyshev nodes initialized by \texttt{Cheb\_init\_cont\_xA}.}. Assume we are interested in the grid point \texttt{i}. Its grid coordinates are stored in the array \texttt{*ind} by calling the function \texttt{void index(int *ind, int *n, int dim, int i)}, where \texttt{*n} are the number of grid points in every external momentum, given by \texttt{n\_A}, and \texttt{dim} is the number of external momenta, 
given by \texttt{dim\_x}. Then we can obtain the index of the \texttt{idir}-th component of the \texttt{i}-th grid point with the function \texttt{int xA\_index(int *ind, int idir, void *dse\_param)}.
When \texttt{x\_A} is correctly initialized access to the coordinates of the \texttt{i}-th grid point is provided via the function \texttt{void outer\_argument(int i, void *dse\_param)} which stores them in \texttt{outer\_arguments}.

\subsection{The array \texttt{A} for linear interpolation}

When linear interpolation is used the array \texttt{A} contains the function values at the external momenta \texttt{x\_A}. The function \texttt{int A\_index(int *ind, int idress, void *dse\_param)} returns the index of the grid point with the coordinates \texttt{ind} obtained with the help of the function \texttt{index}, see section \ref{sec:x_A}. The argument \texttt{idress} is the index of the dressing function.

\subsection{The array \texttt{A} for Chebyshev expansion}

For a Chebyshev expansion the array \texttt{A} contains the Chebyshev coefficients. They can be initialized from a (user-provided) function \texttt{func} by calling \texttt{void Cheb\_init\-\_coeff\_mult(double (*func)(double} \texttt{*x, void} \texttt{*param),} \texttt{ void} \texttt{(*traf)(double} \texttt{*x,} \texttt{void *param), int} \texttt{dim, int} \texttt{*n\_f,} \texttt{int}\texttt{ *n\_c,} \texttt{double} \texttt{*c\_ar,} \texttt{void} \texttt{*param)}. The argument \texttt{traf} defines the transformation of the domain of the Chebyshev polynomials ($[-1,1]$ and direct products thereof) to whatever is the domain of the function. We recommend the function \texttt{void Cheb\_gen\_traf(double *x, void *param)} for this purpose which transforms the interval depending on \texttt{int *cheb\-\_trafo}, a member of the corresponding \texttt{dse} structure. If \texttt{cheb\_trafo[i]} is \texttt{0/1/2} no transformation/a linear transformation/a logarithmic transformation is employed for the \texttt{i}-th external momentum. In what follows we give the expressions usually used as arguments of \texttt{Cheb\_init\_coeff\_mult} in parenthesis. \texttt{dim} defines the number of continuous arguments (\texttt{dim\_x - dim\_mat}),
\texttt{*n\_f} (\texttt{n\_A}) is the number of points used for evaluating the inner product to calculate the \texttt{*n\_c} (\texttt{n\_A}) Chebyshev coefficients stored in \texttt{*c\_ar} (\texttt{A}) and \texttt{param} are some parameters (\texttt{dse\_param}). In general \texttt{n\_f}=\texttt{n\_c} should be used. We note here that if more than one dressing function is to be interpolated the Chebyshev coefficients can be stored in one array \texttt{A} by simply passing the argument \texttt{A+idress*ntot\_A} as \texttt{*c\_ar} where \texttt{idress} denotes the \texttt{idress}-th function.

\section{Structure of user-defined functions}
\label{sec:app_extra_user-funcs}

For the following members of \texttt{dse} or \texttt{quad} structures the user has to provide functions:

\begin{itemize}
 \item \texttt{void def\_domain(double *x, void *dse\_param)}: Here the user defines the boundaries of the interpolation domain for the dressing functions. The bounds are saved in the array \texttt{double *domain}, a member of the structure \texttt{dse\_param}. The lower and upper bounds of the \texttt{i}-th external momentum are saved in \texttt{domain[2*i]} and \texttt{domain[2*i+1]}, respectively. The array \texttt{x} refers to the external variables. As examples consider the following functions: \texttt{scalgluevert\-\_def\_domain} in \textit{scalar\_QCD.cpp} or \texttt{interp\_def\_domain} in \textit{interp.cpp}. \texttt{def\-\_domain} is a member of a \texttt{dse} structure.

 \item \texttt{double} \texttt{interp\_offdomain(int} \texttt{*pos,} \texttt{double} \texttt{*x,} \texttt{int} \texttt{iA,} \texttt{void} \texttt{*dse\_param)}: This function is required for extrapolation, i.~e., when the domain of the interpolation of a dressing function is left. The array \texttt{pos} of length \texttt{dim\_x} contains the values \texttt{0}, \texttt{1} and \texttt{2} for every external momentum. If \texttt{pos[i]=0}, the interpolation domain for the \texttt{i}-th external momentum is not left, while a value of \texttt{1} or \texttt{2} means that the lower or upper bounds are crossed, respectively. \texttt{pos} is created automatically based on the information provided in \texttt{def\_domain}. \texttt{iA} tells which of the \texttt{dim\_A} dressing functions is required and the array \texttt{x} contains the values of the external momenta. \texttt{dse\_param} refers to the \texttt{dse} structure to which this dressing function belongs. Examples are the functions \texttt{ghprop\_interp\_offdomain\_cheb} in \textit{YM4d.cpp} or \texttt{scalvert\_interp\_offdomain} in \textit{scalar\_QCD.cpp}. \texttt{interp\_offdomain} is a member of a \texttt{dse} structure.

 \item \texttt{void (*renorm)(void *dse\_param)}: This function has to perform several tasks. It does not only implement the renormalization procedure but also contains all steps required to proceed from the results of the integration to the expressions required by the solving algorithms.

 The self-energy contributions of single diagrams are calculated by the solving algorithms with the function \texttt{selfenergy}. It stores the results for each dressing function and diagram in the \texttt{dse} member \texttt{double *self\_A}. This array is organized as follows: For every quadrature \texttt{Q[i]} a block of size \texttt{n\_loop[i]*dim\_A*ntot\_A} is used. These blocks are separated into \texttt{ntot\_A} sub-blocks of size \texttt{dim\_A*n\_loop[i]}, i.~e., each entry contains the result of the integration of one diagram for a specific external momentum. Appropriately summed up we obtain the result for the right-hand side of a DSE for all external momenta. This summation as well as the renormalization have to be done in \texttt{renorm}. Furthermore, if the DSE is not projected directly onto its dressing functions, the linear system of equations to obtain them has to be solved here. Finally, depending on the solving algorithm for the DSEs, either the \texttt{dse} members \texttt{A} or \texttt{E} have to be set: In the case of a fixed point iteration the results can be directly saved to \texttt{A} if a linear interpolation is used. For a Chebyshev expansion the new coefficients \texttt{A} are calculated with \texttt{Cheb\_coeff\_mult}. If the solving algorithm is Newton's method the renormalization function calculates $E^{(i,k)}$ of \eref{eq:E}. Examples are \texttt{scalgluevert\_renorm} in \textit{scalar\_QCD.cpp} and \texttt{prop\_renorm\_cheb\_secant} in \textit{YM4d.cpp}. \texttt{renorm} is a member of a \texttt{dse} structure.

 \item \texttt{void integrand(double *erg, double *x, void *int\_param)}: The actual kernels for the integration are contained in this function. Every \texttt{quad} structure handles the integration of one or several integrals and the values of the integrands at the integration variables given by the array \texttt{x} are stored in the array \texttt{erg}. The external momenta are accessed via \texttt{int\_param}, which refers to a \texttt{dse} structure. Its member \texttt{double *outer\_arguments} has to contain the external momenta. \texttt{integrand} can be a user-written function, but more commonly it will be created by the \textit{Mathematica} functions of \textit{CrasyDSE}. Examples are \texttt{sphere\_integrand} in \textit{sphere.cpp}, which was created manually and is thus relatively simple, and \texttt{kernelsVertex} in \textit{kernelsAll.cpp} of the scalar QCD example, which was created in \textit{CrasyDSE\_YM+scalar.nb}. \texttt{integrand} is a member of a \texttt{quad} structure.

 \item \texttt{double jacob(double *x)}: This function can be used for the Jacobian of the integral measure. Since the Jacobian is automatically taken into account for every integration, it must always be defined. Thus, if it is already contained in the integrand, this function must return $1$. The array \texttt{x} holds the integration variables. Examples are \texttt{sphere\_jacob} in \textit{sphere.cpp} and \texttt{ghprop\_jacob} in \textit{YM4d.cpp}. \texttt{jacob} is a member of a \texttt{quad} structure.

 \item \texttt{void coeff(double *coefficients, void *int\_param)}: Any trivial coefficients of the integrals which do not depend on any momenta can be put into this function. Their values are saved in the array \texttt{coefficients} which is multiplied with the results from the integration. \texttt{int\_param} refers to the \texttt{quad} structure of the integration. Similar to \texttt{integrand} this function can be written manually or created with \textit{Mathematica}. Examples are \texttt{sphere\_coeff} in \textit{sphere.cpp}, which was created manually, and \texttt{coeffsVertex} in \textit{kernelsAll.cpp} of the scalar QCD example, which was created in \textit{CrasyDSE\_YM+scalar.nb}. \texttt{coeff} is a member of a \texttt{quad} structure.

 \item \texttt{void} \texttt{boundary(double} \texttt{*bound,} \texttt{double} \texttt{*x,} \texttt{int} \texttt{idim,} \texttt{void} \texttt{*int\_param)}: The limits of the integration are defined in this function. For the \texttt{idim}-th integration variable \texttt{bound[0]} and \texttt{bound[1]} are set to the lower and upper integration bounds, respectively. The array \texttt{x} contains the external momenta, on which the integration bounds may depend. If this is the case, the variable \texttt{bound\_type} has to be set to \texttt{1}, see section~\ref{sec:ImplementationC_quad}. \texttt{int\_param} refers to the \texttt{quad} structure of the integration. Examples are \texttt{sphere\_boundary} in \textit{sphere.cpp} and \texttt{scalgluevert\_boundary} in \textit{scalar\_QCD.cpp}. \texttt{boundary} is a member of a \texttt{quad} structure.

\end{itemize}

\section{Overview of all \textit{C++} functions}

This appendix provides tables with all \textit{C++} functions of \textit{CrasyDSE} relevant for the user. Table~\ref{tab:functions} contains the most prominent functions of \textit{CrasyDSE}, table~\ref{tab:approx} the functions and variables for approximating the dressing functions, table~\ref{tab:quad} the functions and variables for the integration and table~\ref{tab:dse} further functions and variables of \texttt{dse} structures. Details on the functions' arguments are provided in the code in the function descriptions. For the user's convenience we provide the file \textit{template\_DSE.cpp} in the directory \textit{main} which contains a list of all functions and variables that have to be defined for a \texttt{dse} structure.

\FloatBarrier

\begin{longtable}{ | p{6.0cm} | p{3.0cm} | p{5.0cm} |}
  \caption{\label{tab:functions} Overview of the main \textit{C++} functions of \textit{CrasyDSE}.}\\
    \hline
    \textbf{Function} &  \textbf{File} & \textbf{Short description} \\ \hline
    \endfirsthead
    \hline
    \textbf{Function} &  \textbf{File} & \textbf{Short description} \\ \hline
    \endhead
    \textbf{Function} &  \textbf{File} & \textbf{Short description} \\ \hline
    \endfoot
    \textbf{Function} &  \textbf{File} & \textbf{Short description} \\ \hline
    \endlastfoot
    \texttt{int \newline A\_index\newline (int *ind, int dress, void *dse\_param)} & \textit{DSE.cpp} & index of coefficient with coordinates \texttt{ind} of \texttt{dress}-th dressing function\\ \hline
    \texttt{void \newline Cheb\_coeff\_mult \newline(int dim, int *n\_f, int *n\_c, double *f\_ar, double *c\_ar)} & \textit{dressing.cpp} & Chebyshev coefficients from array of function values at Chebyshev nodes\\ \hline
    \texttt{double \newline Cheb\_gen\_dress\_interp\newline (double *x, int i, void *dse\_param)} & \textit{dressing.cpp} & interpolating Chebyshev polynomial \\ \hline
    \texttt{void \newline Cheb\_gen\_traf\newline (double *x, void *param)} & \textit{dressing.cpp} & maps domain of Chebyshev polynomials to domain of function\\ \hline
    \texttt{void \newline Cheb\_init\_coeff\_mult\newline (double (*func)(double *x, void *param), void (*traf)(double *x, void *param), int dim, int *n\_f, int *n\_c, double *c\_ar, void *param)} & \textit{dressing.cpp} & Chebyshev coefficients from function\\ \hline
    \texttt{void \newline Cheb\_init\_cont\_xA\newline (void *dse\_param)} & \textit{dressing.cpp} & initializes the continuous external variables to the Chebyshev nodes\\ \hline
    \texttt{void \newline dealloc\_A\_xA\newline (void *dse\_param)} & \textit{DSE.cpp} & deallocate members of \texttt{dse} structure\\ \hline
    \texttt{void \newline dealloc\_quad\newline (void *quad\_param)} & \textit{quadrature.cpp} & deallocate members of \texttt{quad} structure\\ \hline
    \texttt{void \newline get\_nw\newline (double *x, double *w, int i, int idim, void *bound\_params, void *quad\_param)} & \textit{quadrature.cpp} & transformed nodes and weights\\ \hline
    \texttt{void \newline index\newline (int *i, int *n, int dim, int index)} & \textit{function.cpp} & coordinates of \texttt{index}-th entry of a one-dimensional array in a \texttt{n[0]}$\times \ldots\times$\texttt{n[dim-1]} grid \\ \hline
    \texttt{void \newline init\_A\_xA\newline (void *dse\_param)} & \textit{DSE.cpp} & allocate  members of \texttt{dse} structure\\ \hline
    \texttt{void \newline init\_dse\_default\newline(void *dse\_param)} & \textit{DSE.cpp} & default values of parameters for allocation of \texttt{dse} structure\\ \hline
    \texttt{void \newline init\_quad\newline (void *quad\_param)} & \textit{quadrature.cpp} & allocate members of \texttt{quad} structure\\ \hline
    \texttt{void \newline integrate\newline (double *erg, void *quad\_param, void *int\_param)} & \textit{quadrature.cpp} & integrate a set of functions\\ \hline
    \texttt{double \newline Lin\_gen\_dress\_interp\newline(double *x, int i, void *dse\_param)} & \textit{dressing.cpp} & linear interpolation of array of function values\\ \hline
    \texttt{void \newline meta\_solve\_iter\newline(struct dse *DSE, int ndse, double epsabsstop, double epsrelstop, int iterstop, int output)} & \textit{DSE.cpp} & solves coupled set of DSEs by iteration\\ \hline
    \texttt{void \newline outer\_argument\newline(int i, void *dse\_param)} & \textit{DSE.cpp} & sets \texttt{outer\_arguments} to the \texttt{i}-th external momenta\\ \hline
    \texttt{void \newline plot\_credits()} & \textit{DSE.cpp} & prints version and author information\\ \hline
    \texttt{void \newline plot\_dressing\newline(void *dse\_param)} & \textit{dressing.cpp} & output of dressing functions and interpolation points on screen\\ \hline
    \texttt{void \newline solve\_iter\newline(void *dse\_param, int output)} & \textit{DSE.cpp} & solves one DSE by iteration\\ \hline
    \texttt{void \newline solve\_iter\_secant\newline(struct dse *DSE, int ndse, int maxiter, double eps\_E, int output)} & \textit{DSE.cpp} & solves array of DSEs with Newton's method\\ \hline
    \texttt{void \newline write\_dressing\newline(void *dse\_param, char *name, double *addpara, int length)} & \textit{dressing.cpp} & output of dressing functions and interpolation points in file\\ \hline
    \texttt{int \newline xA\_index\newline(int *ind, int idir, void *dse\_param)} & \textit{DSE.cpp} & index of \texttt{idir}-th direction of a grid point with coordinates \texttt{ind}\\ \hline
\end{longtable}

\begin{longtable}{ | p{4.0cm} | p{5.0cm} | p{5.0cm} |}
  \caption{\label{tab:approx} User defined members of a \texttt{dse} structure relevant for the approximation of functions. References to the scalar-gauge field vertex refer to the example of section \ref{sec:example}, those to the ghost-gluon system to the example of section \ref{sec:YM}.}\\
    \hline
    \textbf{Member} &  \textbf{General purpose} & \textbf{DSE usage/example} \\ \hline
    \endfirsthead
    \hline
    \textbf{Member} &  \textbf{General purpose} & \textbf{DSE usage/example} \\ \hline
    \endhead
    \textbf{Member} &  \textbf{General purpose} & \textbf{DSE usage/example} \\ \hline
    \endfoot
    \textbf{Member} &  \textbf{General purpose} & \textbf{DSE usage/example} \\ \hline
    \endlastfoot
    \texttt{double \newline *A} & interpolation coefficients & e.g., coefficients of $A_{p_1}$ and $A_{p_2}$ for the scalar-gauge field vertex\\ \hline
    \texttt{int \newline *cheb\_func\_trafo} & defines transformation of approximated function for Chebyshev expansion; \texttt{0}: none, \texttt{1}: logarithmic & e.g., approximation of the exponential of the ghost dressing function instead of the dressing function itself \\ \hline    
    \texttt{int \newline *cheb\_trafo} & defines transformation function to interval $[-1,1]$ for Chebyshev expansion; \texttt{0}: none, \texttt{1}: linear, \texttt{2}: logarithmic & e.g., logarithmic transformation of definition interval of ghost dressing function\\ \hline
    \texttt{double \newline *cheb\_x} & necessary for Chebyshev interpolation with \texttt{dim\_mat}$>0$ & see \textit{interp.cpp} for correct usage \\ \hline
    \texttt{void \newline (*def\_domain)\newline (double *x, void *dse\_param)} & defines domain of definition & e.g., $\left[0,\left(\Lambda/2\right)^2\right]^2\times\left[-1,1\right]$ for scalar-gauge field vertex \\ \hline
    \texttt{int \newline dim\_A} & number of interpolated functions & number of dressing functions for a given Green function, e.g., $2$ ($A_{p_1}$ and $A_{p_2}$) for the scalar-gauge field vertex\\ \hline
    \texttt{int \newline dim\_mat} & number of discrete arguments of interpolated functions & number of independent Matsubara frequencies\\ \hline
    \texttt{int \newline dim\_x} & number of arguments of interpolated functions & number of arguments of dressing functions, e.g., $3$ ($p_1^2,p_2^2$ and $z$) for the scalar-gauge field vertex\\ \hline
    \texttt{double \newline (*dress)\newline(double *x, int i, void *dse\_param)} & returns value of the \texttt{i}-th dressing function at the momenta \texttt{x} & e.g., $A_{p_1}$ and $A_{p_2}$ for the scalar-gauge field vertex\\ \hline    
    \texttt{double \newline *E} & array of the values on the left-hand side of \eref{eq:E} for Newton's method & has to be calculated in the renormalization procedure defined by the user, e.g., in     \texttt{prop\_renorm\_cheb\_secant} for the ghost-gluon system\\ \hline    
    \texttt{void \newline (*init\_func)\newline (double *x, void *dse\_param)} & function used for initialization of the Chebyshev coefficients & ansatz for dressing function with the transformation from \texttt{cheb\_func\_trafo} taken into account, e.g., ghost dressing function ansatz given by \texttt{ansatzGh4dLog} \\ \hline    
    \texttt{double \newline (*interp\_offdomain)\newline (int *pos, double *x, int iA, void *dse\_param)} & is called if \texttt{dress} is evaluated outside domain & e.g., $A_{p_1,p_2}=\hat{Z}_1$ for $p_1,p_2\geq\left(\Lambda/2\right)^2$ in scalar-gauge field vertex \\ \hline
    \texttt{int \newline *n\_A} & numbers of expansion coefficients in the \texttt{dim\_x} variables & numbers of expansion coefficients for every external momentum\\ \hline    
    \texttt{double \newline *x\_A} & (discrete) interpolation points & external momenta (Matsubara frequencies)\\ \hline 
\end{longtable}

\begin{longtable}{ | p{4.0cm} | p{5.0cm} | p{5.0cm} |}
  \caption{\label{tab:quad}User defined members of a \texttt{quad} structure. References to the scalar-gauge field vertex refer to the example of section \ref{sec:example}, those to the ghost-gluon system to the example of section \ref{sec:YM}.}\\
    \hline
    \textbf{Member} &  \textbf{General purpose} & \textbf{DSE usage/example} \\ \hline
    \endfirsthead
    \hline
    \textbf{Member} &  \textbf{General purpose} & \textbf{DSE usage/example} \\ \hline
    \endhead
    \textbf{Member} &  \textbf{General purpose} & \textbf{DSE usage/example} \\ \hline
    \endfoot
    \textbf{Member} &  \textbf{General purpose} & \textbf{DSE usage/example} \\ \hline
    \endlastfoot
    \texttt{int\newline bound\_type} & defines if the integration boundaries depend on the \texttt{nint\_para} sets of parameters & defines if the integration boundaries depend on the external momenta\\ \hline
    \texttt{void \newline (*boundary)\newline (double *bound, double *x, int idim, void *int\_param)} & defines the integration boundaries for the different integration regions & e.g., for the scalar-gauge field vertex \texttt{qs}$\in[0,\Lambda^2]$, \texttt{ct1}, \texttt{ct2} $\in[-1,1]$  \\ \hline
    \texttt{void\newline (*coeff)\newline (double *coefficients, void *int\_param)} & constant factor of integrand & constant factors of self-energy kernels\\ \hline
    \texttt{int \newline dim} & number of integration variables & number of loop momentum variables, e.g., $3$  (\texttt{qs}, \texttt{ct1} and \texttt{ct2}) for scalar-gauge field vertex \\ \hline
    \texttt{void \newline (*init\_para)\newline (int i, void *int\_param)} & initializes the \texttt{nint\_para} parameter values  & automatically set to void \texttt{outer\_argument} which initializes the external momenta  \\ \hline
    \texttt{void\newline (*integrand)\newline (double *erg, double *x, void *int\_param)} & the integrand & kernels of the self-energies \\ \hline
    \texttt{double\newline (*jacob)\newline (double *x)} & Jacobian & usually set to one\\ \hline
    \texttt{int \newline nint} & number of different integrands that are integrated over the same variables  & number of dressing functions multiplied by number of graphs with the same integration variables, e.g.,  $2*2=4$ for the scalar-gauge field vertex \\ \hline
    \texttt{int \newline nint\_para} & number of times the integrands are integrated for different parameter values  & number of external momenta at which the self-energy is evaluated, i.e., number of interpolation points \\ \hline
    \texttt{int \newline *n\_part} & number of integration regions for any of the \texttt{dim} integration variables & e.g., the radial integration for the ghost-gluon system is split into two parts \\ \hline
    \texttt{int \newline *nodes\_part} & number of quadrature nodes for the different integration regions  & e.g., the radial integration of the ghost-gluon system \\ \hline
    \texttt{double \newline *param} & defines additional parameters in a quadrature rule, e.g., for the implemented double exponential & not used in any of the DSE example, correct usage demonstrated in \textit{sphere\_main.cpp}\\ \hline  
    \texttt{int \newline *traf} & defines the transformation function from $[-1,1]$ to the true region of integration for different integration regions & e.g., for the scalar-gauge field vertex the \texttt{qs} integration is mapped via a modified logarithmic mapping whereas the integrations in \texttt{ct1} and \texttt{ct2} are not mapped\\ \hline
    \texttt{int \newline *type} & quadrature rules for the different integration regions  & e.g., the scalar gluon vertex uses Fejer's second rule for the \texttt{qs} integration, Gauss-Chebyshev quadrature for the \texttt{ct1} integration and Gauss-Legendre quadrature for the \texttt{ct2} integration\\ \hline
\end{longtable}

\begin{longtable}{ | p{4.0cm}  | p{10.0cm} |}
  \caption{\label{tab:dse} User defined members of a \texttt{dse} structure relevant for solving DSEs (not already included in table~\ref{tab:approx}).}\\
    \hline
    \textbf{Member} &  \textbf{Short description} \\ \hline
    \endfirsthead
    \hline
    \textbf{Member} &  \textbf{Short description} \\ \hline
    \endhead
    \textbf{Member} &  \textbf{Short description} \\ \hline
    \endfoot
    \textbf{Member} &  \textbf{Short description} \\ \hline
    \endlastfoot
    \texttt{double \newline anom\_dim} & anomalous dimension of a dressing function; optional\\ \hline
    \texttt{std::string \newline DSE\_name} & name of the DSE; optional\\ \hline
    \texttt{double \newline epsabs} & stopping criterion for iteration: absolute difference of solutions\\ \hline
    \texttt{double \newline epsrel} & stopping criterion for iteration: relative difference of solutions\\ \hline
    \texttt{double \newline h} & stepsize for Broyden's method; required only for secant method\\ \hline
    \texttt{void \newline (*init\_dress)\newline (void *dse\_param)} & initializes coefficients \texttt{*n\_A} and interpolation points, i.e., initial guess for the dressing function and definition of external momenta\\ \hline    
    \texttt{void \newline (*init\_renormparam)\newline (void *dse\_param)} & initializes renormalization parameters\\ \hline
    \texttt{int \newline it\_counter} & counter for iterations; optional\\ \hline
    \texttt{int \newline maxiter} & stopping criterion for iteration: maximal iteration number\\ \hline
    \texttt{void \newline (*mod)} & model parameters, usually the same for all Green functions\\ \hline
    \texttt{int \newline *n\_loop} & number of diagrams with the same integral, \texttt{*n\_loop[i]}, \texttt{i}$\in\{0,\dots,$\texttt{n\_loopnumber}$-1\}$\\ \hline    
    \texttt{int \newline n\_loopnumber} & number of different integrals\\ \hline    
    \texttt{int \newline n\_otherGF} & number of other Green functions contributing to the given DSE\\ \hline    
    \texttt{struct dse \newline *otherGF} & one \texttt{dse} structure for every other Green function contributing\\ \hline    
    \texttt{double \newline *outer\_arguments} & stores the external momentum for a given index when the function \texttt{outer\_argument} is called\\ \hline    
    \texttt{struct quad \newline *Q} & one \texttt{quad} structure for every of the \texttt{n\_loopnumber} different integrals\\ \hline    
    \texttt{void \newline (*renorm)\newline (void *dse\_param)} & function that is called after calculating the loop integrals of the self-energy in iteration based solving routines\\ \hline
    \texttt{int \newline *renorm\_i} & index of renormalization point\\ \hline
    \texttt{int \newline renorm\_init} & determines if \texttt{*Z\_renorm}, \texttt{*renorm\_param}, \texttt{*renorm\_x} and \texttt{*renorm\_i} are allocated automatically by \texttt{init\_A\_xA}\\ \hline
    \texttt{int \newline renorm\_n\_param} & number of renormalization parameters\\ \hline
    \texttt{int \newline renorm\_n\_Z} & number of renormalization constants\\ \hline
    \texttt{double \newline *renorm\_param} & stores renormalization parameters\\ \hline
    \texttt{double \newline *renorm\_x} & renormalization point\\ \hline
    \texttt{double \newline *Z\_renorm} & stores renormalization constants\\ \hline
\end{longtable}

\FloatBarrier

\bibliographystyle{utphys_mod}
\bibliography{literature_solveDSEs}

\providecommand{\href}[2]{#2}\begingroup\raggedright\begin{thebibliography}{10}

\bibitem{Berges:2000ew}
J.~Berges, N.~Tetradis, and C.~Wetterich, {\em Phys. Rept.} {\bf 363} (2002)
  223--386,
\href{http://arxiv.org/abs/hep-ph/0005122}{{\tt arXiv:hep-ph/0005122}}.

\bibitem{Pawlowski:2005xe}
J.~M. Pawlowski, \href{http://dx.doi.org/10.1016/j.aop.2007.01.007}{{\em Annals
  Phys.} {\bf 322} (2007)  2831--2915},
\href{http://arxiv.org/abs/hep-th/0512261}{{\tt arXiv:hep-th/0512261}}.

\bibitem{Gies:2006wv}
H.~Gies, \href{http://arxiv.org/abs/hep-ph/0611146}{{\tt
  arXiv:hep-ph/0611146}}.
Presented at ECT* School on Renormalization Group and Effective Field Theory
  Approaches to Many-Body Systems, Trento, Italy, 27 Feb - 10 Mar 2006.

\bibitem{Rosten:2010vm}
O.~J. Rosten, \href{http://dx.doi.org/10.1016/j.physrep.2011.12.003}{{\em
  Phys.Rept.} {\bf 511} (2012)  177--272},
\href{http://arxiv.org/abs/1003.1366}{{\tt arXiv:1003.1366 [hep-th]}}.

\bibitem{Alkofer:2000wg}
R.~Alkofer and L.~von Smekal, {\em Phys. Rept.} {\bf 353} (2001)  281,
\href{http://arxiv.org/abs/hep-ph/0007355}{{\tt arXiv:hep-ph/0007355}}.

\bibitem{Alkofer:2008nt}
R.~Alkofer, M.~Q. Huber, and K.~Schwenzer,
  \href{http://dx.doi.org/10.1016/j.cpc.2008.12.009}{{\em Comput. Phys.
  Commun.} {\bf 180} (2009)  965--976},
\href{http://arxiv.org/abs/0808.2939}{{\tt arXiv:0808.2939 [hep-th]}}.

\bibitem{Binosi:2009qm}
D.~Binosi and J.~Papavassiliou,
  \href{http://dx.doi.org/10.1016/j.physrep.2009.05.001}{{\em Phys. Rept.} {\bf
  479} (2009)  1--152},
\href{http://arxiv.org/abs/0909.2536}{{\tt arXiv:0909.2536 [hep-ph]}}.

\bibitem{Roberts:2012sv}
C.~D. Roberts,
\href{http://arxiv.org/abs/1203.5341}{{\tt arXiv:1203.5341 [nucl-th]}}.

\bibitem{Berges:2004pu}
J.~Berges, \href{http://dx.doi.org/10.1103/PhysRevD.70.105010}{{\em Phys. Rev.}
  {\bf D70} (2004)  105010},
\href{http://arxiv.org/abs/hep-ph/0401172}{{\tt arXiv:hep-ph/0401172}}.

\bibitem{vonSmekal:1997vx}
L.~von Smekal, A.~Hauck, and R.~Alkofer,
  \href{http://dx.doi.org/10.1006/aphy.1998.5806}{{\em Ann. Phys.} {\bf 267}
  (1998)  1},
\href{http://arxiv.org/abs/hep-ph/9707327}{{\tt arXiv:hep-ph/9707327}}.

\bibitem{Fischer:2003rp}
C.~S. Fischer and R.~Alkofer,
  \href{http://dx.doi.org/10.1103/PhysRevD.67.094020}{{\em Phys. Rev.} {\bf
  D67} (2003)  094020},
\href{http://arxiv.org/abs/hep-ph/0301094}{{\tt arXiv:hep-ph/0301094}}.

\bibitem{Fischer:2008uz}
C.~S. Fischer, A.~Maas, and J.~M. Pawlowski,
  \href{http://dx.doi.org/10.1016/j.aop.2009.07.009}{{\em Annals Phys.} {\bf
  324} (2009)  2408--2437},
\href{http://arxiv.org/abs/0810.1987}{{\tt arXiv:0810.1987 [hep-ph]}}.

\bibitem{Huber:2009wh}
M.~Q. Huber, K.~Schwenzer, and R.~Alkofer,
  \href{http://dx.doi.org/10.1140/epjc/s10052-010-1371-x}{{\em Eur. Phys. J.}
  {\bf C68} (2010)  581--600},
\href{http://arxiv.org/abs/0904.1873}{{\tt arXiv:0904.1873 [hep-th]}}.

\bibitem{vonSmekal:1997is}
L.~von Smekal, R.~Alkofer, and A.~Hauck,
  \href{http://dx.doi.org/10.1103/PhysRevLett.79.3591}{{\em Phys. Rev. Lett.}
  {\bf 79} (1997)  3591--3594},
\href{http://arxiv.org/abs/hep-ph/9705242}{{\tt arXiv:hep-ph/9705242}}.

\bibitem{Hauck:1998fz}
A.~Hauck, L.~von Smekal, and R.~Alkofer,
  \href{http://dx.doi.org/10.1016/S0010-4655(98)00045-9}{{\em Comput. Phys.
  Commun.} {\bf 112} (1998)  166},
\href{http://arxiv.org/abs/hep-ph/9804376}{{\tt arXiv:hep-ph/9804376}}.

\bibitem{Wolfram:2004}
S.~Wolfram, {\em The Mathematica Book}. Wolfram Media and Cambridge University
  Press, 2004.

\bibitem{Huber:2011qr}
M.~Q. Huber and J.~Braun,
  \href{http://dx.doi.org/10.1016/j.cpc.2012.01.014}{{\em Comput.Phys.Commun.}
  {\bf 183} (2012)  1290--1320},
\href{http://arxiv.org/abs/1102.5307}{{\tt arXiv:1102.5307 [hep-th]}}.

\bibitem{Huber:2012th}
M.~Q. Huber, {\em {On gauge fixing aspects of the infrared behavior of
  Yang-Mills Green functions}}. Springer, 2012.
\newblock \href{http://arxiv.org/abs/1005.1775}{{\tt arXiv:1005.1775}}.
\newblock Ph.D.Thesis, Karl-Franzens-University Graz, 2010.

\bibitem{Huber:2009tx}
M.~Q. Huber, R.~Alkofer, and S.~P. Sorella,
  \href{http://dx.doi.org/10.1103/PhysRevD.81.065003}{{\em Phys. Rev.} {\bf
  D81} (2010)  065003},
\href{http://arxiv.org/abs/0910.5604}{{\tt arXiv:0910.5604 [hep-th]}}.

\bibitem{Huber:2010cq}
M.~Q. Huber, R.~Alkofer, and S.~P. Sorella,
  \href{http://dx.doi.org/10.1063/1.3574962}{{\em AIP Conf.Proc.} {\bf 1343}
  (2011)  158--160},
\href{http://arxiv.org/abs/1010.4802}{{\tt arXiv:1010.4802 [hep-th]}}.

\bibitem{Huber:2011fw}
M.~Q. Huber, R.~Alkofer, and K.~Schwenzer, {\em PoS} {\bf FACESQCD} (2010)
  001,
\href{http://arxiv.org/abs/1103.0236}{{\tt arXiv:1103.0236 [hep-th]}}.

\bibitem{Fischbacher:2012ib}
T.~Fischbacher and F.~Synatschke-Czerwonka,
\href{http://arxiv.org/abs/1202.5984}{{\tt arXiv:1202.5984 [physics.comp-ph]}}.

\bibitem{Bogoliubov:1957gp}
N.~Bogoliubov and O.~Parasiuk, {\em Acta Math.} {\bf 97} (1957)  227--266.

\bibitem{Zimmermann:1969jj}
W.~Zimmermann, {\em Commun.Math.Phys.} {\bf 15} (1969)  208--234.

\bibitem{Hepp:1966eg}
K.~Hepp, {\em Commun.Math.Phys.} {\bf 2} (1966)  301--326.

\bibitem{Bloch:1995dd}
J.~C. Bloch, \href{http://arxiv.org/abs/hep-ph/0208074}{{\tt
  arXiv:hep-ph/0208074 [hep-ph]}}.
Ph.D. thesis (1995), University of Durham.

\bibitem{Atkinson:1997tu}
D.~Atkinson and J.~C.~R. Bloch,
  \href{http://dx.doi.org/10.1103/PhysRevD.58.094036}{{\em Phys. Rev.} {\bf
  D58} (1998)  094036},
\href{http://arxiv.org/abs/hep-ph/9712459}{{\tt arXiv:hep-ph/9712459}}.

\bibitem{Maas:2005xh}
A.~Maas, \href{http://dx.doi.org/10.1016/j.cpc.2006.02.005}{{\em Comput. Phys.
  Commun.} {\bf 175} (2006)  167--179},
\href{http://arxiv.org/abs/hep-ph/0504110}{{\tt arXiv:hep-ph/0504110}}.

\bibitem{Horvatic:2011pc}
D.~Horvatic, private communications, 2011.

\bibitem{Alkofer:2011up}
M.~Mitter, M.~Hopfer, B.-J. Schaefer, and R.~Alkofer, in preparation.

\bibitem{Fischer:2002hn}
C.~S. Fischer and R.~Alkofer,
  \href{http://dx.doi.org/10.1016/S0370-2693(02)01809-9}{{\em Phys. Lett.} {\bf
  B536} (2002)  177--184},
\href{http://arxiv.org/abs/hep-ph/0202202}{{\tt arXiv:hep-ph/0202202}}.

\bibitem{Ball:1980ay}
J.~S. Ball and T.-W. Chiu,
{\em Phys. Rev.} {\bf D22} (1980)  2542.

\bibitem{Maris:1997tm}
P.~Maris and C.~D. Roberts,
  \href{http://dx.doi.org/10.1103/PhysRevC.56.3369}{{\em Phys.Rev.} {\bf C56}
  (1997)  3369--3383}, \href{http://arxiv.org/abs/nucl-th/9708029}{{\tt
  arXiv:nucl-th/9708029 [nucl-th]}}.

\bibitem{Maris:1999nt}
P.~Maris and P.~C. Tandy,
  \href{http://dx.doi.org/10.1103/PhysRevC.60.055214}{{\em Phys.Rev.} {\bf C60}
  (1999)  055214}, \href{http://arxiv.org/abs/nucl-th/9905056}{{\tt
  arXiv:nucl-th/9905056 [nucl-th]}}.

\bibitem{Alkofer:2002bp}
R.~Alkofer, P.~Watson, and H.~Weigel,
  \href{http://dx.doi.org/10.1103/PhysRevD.65.094026}{{\em Phys.Rev.} {\bf D65}
  (2002)  094026}, \href{http://arxiv.org/abs/hep-ph/0202053}{{\tt
  arXiv:hep-ph/0202053 [hep-ph]}}.

\bibitem{Fischer:2009gk}
C.~S. Fischer and J.~A. M\"uller,
  \href{http://dx.doi.org/10.1103/PhysRevD.80.074029}{{\em Phys. Rev.} {\bf
  D80} (2009)  074029},
\href{http://arxiv.org/abs/0908.0007}{{\tt arXiv:0908.0007 [hep-ph]}}.

\bibitem{Alkofer:2008tt}
R.~Alkofer, C.~S. Fischer, F.~J. Llanes-Estrada, and K.~Schwenzer,
  \href{http://dx.doi.org/10.1016/j.aop.2008.07.001}{{\em Annals Phys.} {\bf
  324} (2009)  106--172},
\href{http://arxiv.org/abs/0804.3042}{{\tt arXiv:0804.3042 [hep-ph]}}.

\bibitem{Fister:2010yw}
L.~Fister, R.~Alkofer, and K.~Schwenzer, {\em Phys. Lett.} {\bf B688} (2010)
  237--243,
\href{http://arxiv.org/abs/1003.1668}{{\tt arXiv:1003.1668 [hep-th]}}.

\bibitem{Alkofer:2010tq}
R.~Alkofer, L.~Fister, A.~Maas, and V.~Macher,
  \href{http://dx.doi.org/10.1063/1.3574969}{{\em AIP Conf.Proc.} {\bf 1343}
  (2011)  179--181},
\href{http://arxiv.org/abs/1011.5831}{{\tt arXiv:1011.5831 [hep-ph]}}.

\bibitem{Macher:2011ys}
V.~Macher, A.~Maas, and R.~Alkofer,
\href{http://arxiv.org/abs/1106.5381}{{\tt arXiv:1106.5381 [hep-ph]}}.

\bibitem{Hopfer:2011dt}
M.~Hopfer, diploma thesis (2011), Karl-Franzens-University Graz.

\bibitem{Maas:2011yx}
A.~Maas, {\em PoS} {\bf FACESQCD} (2010)  033,
  \href{http://arxiv.org/abs/1102.0901}{{\tt arXiv:1102.0901 [hep-lat]}}.

\bibitem{Marciano:1977su}
W.~J. Marciano and H.~Pagels,
{\em Phys. Rept.} {\bf 36} (1978)  137.

\bibitem{Aguilar:2008xm}
A.~Aguilar, D.~Binosi, and J.~Papavassiliou,
  \href{http://dx.doi.org/10.1103/PhysRevD.78.025010}{{\em Phys.Rev.} {\bf D78}
  (2008)  025010}, \href{http://arxiv.org/abs/0802.1870}{{\tt arXiv:0802.1870
  [hep-ph]}}.

\bibitem{Fischer:2003zc}
C.~S. Fischer, \href{http://arxiv.org/abs/hep-ph/0304233}{{\tt
  arXiv:hep-ph/0304233 [hep-ph]}}.
Ph.D. thesis, Eberhard-Karls-Universit\"at zu T\"ubingen (2003).

\bibitem{Fischer:2009tn}
C.~S. Fischer and J.~M. Pawlowski,
  \href{http://dx.doi.org/10.1103/PhysRevD.80.025023}{{\em Phys. Rev.} {\bf
  D80} (2009)  025023},
\href{http://arxiv.org/abs/0903.2193}{{\tt arXiv:0903.2193 [hep-th]}}.

\bibitem{Cucchieri:2008qm}
A.~Cucchieri, A.~Maas, and T.~Mendes,
  \href{http://dx.doi.org/10.1103/PhysRevD.77.094510}{{\em Phys. Rev.} {\bf
  D77} (2008)  094510},
\href{http://arxiv.org/abs/0803.1798}{{\tt arXiv:0803.1798 [hep-lat]}}.

\bibitem{Lerche:2002ep}
C.~Lerche and L.~von Smekal, {\em Phys. Rev.} {\bf D65} (2002)  125006,
\href{http://arxiv.org/abs/hep-ph/0202194}{{\tt arXiv:hep-ph/0202194}}.

\bibitem{Schleifenbaum:2004id}
W.~Schleifenbaum, A.~Maas, J.~Wambach, and R.~Alkofer, {\em Phys. Rev.} {\bf
  D72} (2005)  014017,
\href{http://arxiv.org/abs/hep-ph/0411052}{{\tt hep-ph/0411052}}.

\bibitem{Pennington:2011xs}
M.~Pennington and D.~Wilson,
  \href{http://dx.doi.org/10.1103/PhysRevD.84.094028}{{\em Phys.Rev.} {\bf D84}
  (2011)  119901},
\href{http://arxiv.org/abs/1109.2117}{{\tt arXiv:1109.2117 [hep-ph]}}.

\bibitem{Boucaud:2008ji}
P.~Boucaud, J.-P. Leroy, A.~L. Yaouanc, J.~Micheli, O.~Pene, and
  J.~Rodriguez-Quintero,
  \href{http://dx.doi.org/10.1088/1126-6708/2008/06/012}{{\em JHEP} {\bf 0806}
  (2008)  012},
\href{http://arxiv.org/abs/0801.2721}{{\tt arXiv:0801.2721 [hep-ph]}}.

\bibitem{Zwanziger:2001kw}
D.~Zwanziger, {\em Phys. Rev.} {\bf D65} (2002)  094039,
\href{http://arxiv.org/abs/hep-th/0109224}{{\tt arXiv:hep-th/0109224}}.

\bibitem{Alkofer:2002ne}
R.~Alkofer, C.~Fischer, and L.~von Smekal, {\em Acta Phys.Slov.} {\bf 52}
  (2002)  191,
\href{http://arxiv.org/abs/hep-ph/0205125}{{\tt arXiv:hep-ph/0205125
  [hep-ph]}}.

\bibitem{vonSmekal:2009ae}
L.~von Smekal, K.~Maltman, and A.~Sternbeck,
  \href{http://dx.doi.org/10.1016/j.physletb.2009.10.030}{{\em Phys.Lett.} {\bf
  B681} (2009)  336--342},
\href{http://arxiv.org/abs/0903.1696}{{\tt arXiv:0903.1696 [hep-ph]}}.

\bibitem{Alkofer:2004it}
R.~Alkofer, C.~S. Fischer, and F.~J. Llanes-Estrada, {\em Phys. Lett.} {\bf
  B611} (2005)  279--288,
\href{http://arxiv.org/abs/hep-th/0412330}{{\tt arXiv:hep-th/0412330}}.

\end{thebibliography}\endgroup

\end{document}